\documentclass[twocolumn]{aastex61}
\usepackage{hyperref}
\usepackage{amsmath,amstext}
\usepackage[T1]{fontenc}
\usepackage[figure,figure*]{hypcap}
\usepackage{graphicx}
\usepackage{multirow,bigdelim}
\usepackage{pstricks}
\usepackage{ulem}
\usepackage{hhline}
\usepackage[percent]{overpic}

\def\sun{$_{\odot}$}

\def\f28{${f}_{2-8{\rm keV}}$}

\def\msun{${\cal M}_{\odot}$}

\def\sun{$_{\odot}$}

\def\ergs{erg s$^{-1}$}
\def\ergscm2{erg s$^{-1}$ cm$^{-2}$}
\def\yr-1{yr$^{-1}$}

\def\asec{\ifmmode^{\prime\prime}\else$^{\prime\prime}$\fi}

\def\spt{$\buildrel{\prime\prime}\over .$}
\def\secspt{$\buildrel{\prime\prime}\over .$}

\def\Chandra{{\it Chandra}}
\def\XMM{{\it XMM-Newton}}

\def\Swift{{\it Swift}}
\def\Nustar{{\it NuSTAR}}

\def\sgra{Sgr~A*}
\def\magnetar{SGR~J1745$-$2900}

\received{Dec 21, 2018}
\revised{July 30, 2019} 
\accepted{October 3, 2019}
\submitjournal{ApJ}

\shorttitle{Sgr A* Bright X-ray Flares}
\shortauthors{Haggard {\it et al.}}

\begin{document}

\title{Chandra Spectral and Timing Analysis of Sgr A*'s Brightest X-Ray Flares}

\correspondingauthor{Daryl Haggard}
\email{daryl.haggard@mcgill.ca}

\author[0000-0001-6803-2138]{Daryl Haggard}
\affil{Department of Physics, McGill University, 3600 University Street, Montr{\'e}al, QC H3A 2T8, Canada}
\affil{McGill Space Institute, McGill University, 3550 University Street, Montr{\'e}al, QC H3A 2A7, Canada}
\affil{CIFAR Azrieli Global Scholar, Gravity \& the Extreme Universe Program, Canadian Institute for Advanced Research, 661 University Avenue, Suite 505, Toronto, ON M5G 1M1, Canada}

\author[0000-0002-3310-1946]{Melania Nynka}
\affil{Department of Physics, McGill University, 3600 University Street, Montr{\'e}al, QC H3A 2T8, Canada}
\affil{McGill Space Institute, McGill University, 3550 University Street, Montr{\'e}al, QC H3A 2A7, Canada}
\affil{MIT Kavli Institute for Astrophysics and Space Research, 77 Massachusetts Avenue, Cambridge, MA 02139, USA}

\author{Brayden Mon}
\affil{Department of Physics, McGill University, 3600 University Street, Montr{\'e}al, QC H3A 2T8, Canada}
\affil{McGill Space Institute, McGill University, 3550 University Street, Montr{\'e}al, QC H3A 2A7, Canada}

\author{Noelia de la Cruz Hernandez}
\affil{Department of Physics, McGill University, 3600 University Street, Montr{\'e}al, QC H3A 2T8, Canada}
\affil{McGill Space Institute, McGill University, 3550 University Street, Montr{\'e}al, QC H3A 2A7, Canada}

\author{Michael Nowak}
\affil{Department of Physics, Washington University, 1 Brookings Drive, St. Louis, MO 63130, USA}

\author[0000-0003-3944-6109]{Craig Heinke}
\affil{Department of Physics, University of Alberta, CCIS 4-183, Edmonton AB T6G 2E1, Canada}

\author{Joseph Neilsen}
\affil{Department of Physics, Villanova University, 800 Lancaster Avenue, Villanova, PA 19085, USA}

\author[0000-0003-3903-0373]{Jason Dexter}
\affil{Max-Planck-Institut f{\"u}r Extraterrestrische Physik, Giessenbachstrasse, D-85748 Garching, Germany}
\affil{JILA and Department of Astrophysical and Planetary Sciences, University of Colorado, Boulder, CO 80309, USA}

\author[0000-0002-5786-186X]{P. Chris Fragile}
\affil{Department of Physics and Astronomy, College of Charleston, Charleston, SC 29424, USA}

\author{Fred Baganoff}
\affil{MIT Kavli Institute for Astrophysics and Space Research, 77 Massachusetts Avenue, Cambridge, MA, 02139, USA}

\author[0000-0003-4056-9982]{Geoffrey C. Bower}
\affil{Academia Sinica Institute of Astronomy and Astrophysics, 645 N. A'ohoku Place, Hilo, HI 96720, USA}

\author{Lia R. Corrales}
\affil{Department of Astronomy, University of Michigan, 1085 S. University, Ann Arbor, MI 48109, USA}

\author{Francesco Coti Zelati}
\affil{Institute of Space Sciences (CSIC), Campus UAB, Carrer de Can Magrans s/n, E-08193 Barcelona, Spain}
\affil{Institut d'Estudis Espacials de Catalunya (IEEC), E-08034 Barcelona, Spain}

\author[0000-0002-0092-3548]{Nathalie Degenaar}
\affil{Anton Pannekoek Institute for Astronomy, University of Amsterdam, Science Park 904, NL-1098 XH Amsterdam, the Netherlands}

\author{Sera Markoff}
\affil{Anton Pannekoek Institute for Astronomy, University of Amsterdam, Science Park 904, NL-1098 XH Amsterdam, the Netherlands}
\affil{Gravitational and Astroparticle Physics Amsterdam, U. Amsterdam, Science Park 904, NL-1098 XH Amsterdam, the Netherlands}

\author[0000-0002-6753-2066]{Mark R. Morris}
\affil{University of California, Los Angeles, CA 90095, USA}

\author[0000-0003-0293-3608]{Gabriele Ponti}
\affil{Osservatorio Astronomico di Brera, Via E. Bianchi 46, I-23807 Merate (LC), Italy}
\affil{Max-Planck-Institut f{\"u}r Extraterrestrische Physik, Giessenbachstrasse, D-85748 Garching, Germany}

\author{Nanda Rea}
\affil{Institute of Space Sciences (CSIC), Campus UAB, Carrer de Can Magrans s/n, E-08193 Barcelona, Spain}
\affil{Institut d'Estudis Espacials de Catalunya (IEEC), E-08034 Barcelona, Spain}

\author{J{\"o}ern Wilms}
\affil{Dr. Karl-Remeis-Sternwarte and Erlangen Centre for Astroparticle Physics, Sternwartstr. 7, D-96049, Bamberg, Germany}

\author{Farhad Yusef-Zadeh}
\affil{Department of Physics and Astronomy and Center for Interdisciplinary Exploration and Research in Astrophysics (CIERA), Northwestern University, Evanston, IL 60208, USA}

\begin{abstract}
We analyze the two brightest \Chandra\ X-ray flares detected from Sagittarius A*, with peak luminosities more than 600$\times$ and 245$\times$ greater than the quiescent X-ray emission. The brightest flare has a distinctive double-peaked morphology --- it lasts 5.7 ks ($\sim 2$ hr), with a rapid rise time of 1500 s and a decay time of 2500 s. The second flare lasts 3.4 ks, with rise and decay times of 1700 s and 1400 s. These luminous flares are significantly harder than quiescence: the first has a power-law spectral index $\Gamma = 2.06\pm 0.14$ and the second has $\Gamma = 2.03\pm 0.27$, compared to $\Gamma = 3.0\pm0.2$ for the quiescent accretion flow. These spectral indices (as well as the flare hardness ratios) are consistent with previously detected \sgra\ flares, suggesting that bright and faint flares arise from similar physical processes. Leveraging the brightest flare's long duration and high signal-to-noise, we search for intraflare variability and detect excess X-ray power at a frequency of $\nu \approx 3$ mHz, but show that it is an instrumental artifact and not of astrophysical origin. We find no other evidence (at the 95\% confidence level) for periodic or quasi-periodic variability in either flares' time series. We also search for nonperiodic excess power but do not find compelling evidence in the power spectrum. Bright flares like these remain our most promising avenue for identifying Sgr A*'s short timescale variability in the X-ray, which may probe the characteristic size scale for the X-ray emission region.
\end{abstract}

\keywords{accretion, accretion disks, black hole physics, Galaxy: center, radiation mechanisms: non-thermal, X-rays: individual: Sgr A*}

\section{Introduction}
\label{sec:intro}

\begin{figure*}[th]
\center{
\includegraphics[trim={5cm 0 5cm 0},clip,scale=0.25]{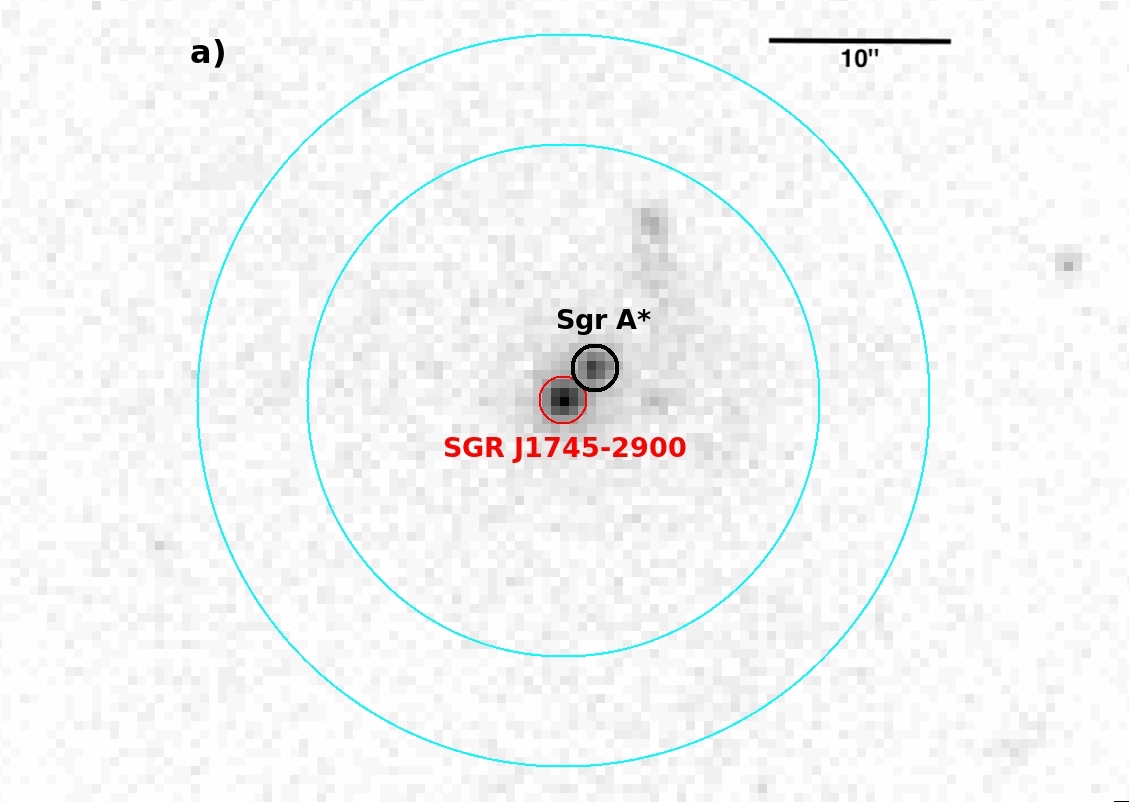}
\includegraphics[scale=0.25]{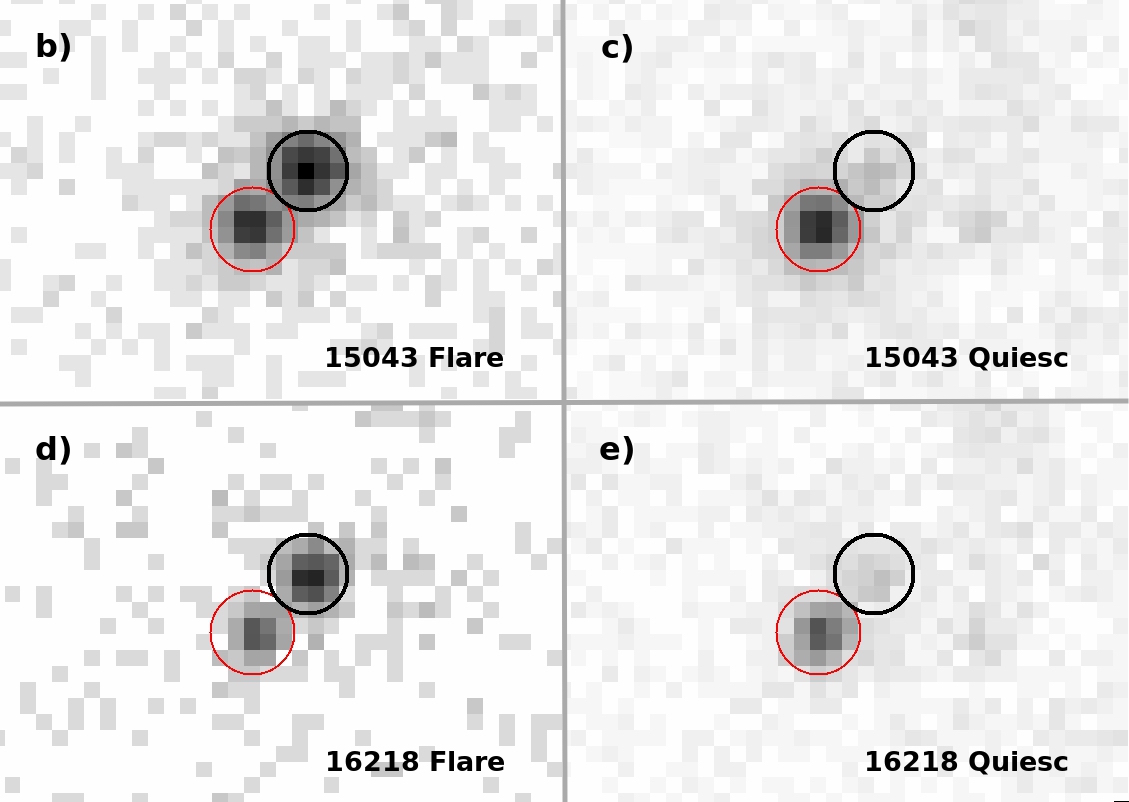}
}
\figcaption{\Chandra\ X-ray images of \sgra\ and the magnetar, \magnetar, on a logarithmic scale in the $2-8$~keV band. Two bright X-ray flares from Sgr A* are clearly visible, as is the decay of the magnetar's flux between 2013 September 14 (ObsID 15043) and 2014 October 20 (ObsID 16218). The large panel (a) shows the full 45.41 ks exposure for ObsID 15043 --- extraction regions for \sgra\ ($\sim$1\spt25 radius) and the magnetar (1\spt3 radius) are shown as black and red circles, respectively, and the background region \citep[matched to the \magnetar\ analysis of][]{CotiZelati17} is marked with a cyan annulus. Smaller panels show the flare and quiescent intervals for ObsID 15043 (b, c) and ObsID 16218 (d, e). The duration of the quiescent periods exceed the flare durations (Table \ref{tab:spec}), so the quiescent images have been rescaled by their exposure time to match the contrast in the flare images.}
\label{fig:fov}
\end{figure*}

Near-infrared observations of massive stars orbiting Sagittarius A* (\sgra\,), the supermassive black hole (SMBH) at the Milky Way's center, have yielded precise measurements of its mass and distance, $(4.28\pm0.1)\times10^6$\msun\ and $7.97\pm0.07$ kpc \citep[e.g.,][]{Ghez08, Gillessen09, Gillessen17, Boehle16, GRAVITY18b, GRAVITY19, Do19}. {\it Chandra} X-ray observations have revealed Sgr A* as a compact ($\sim$1\asec\,), low-luminosity source ($L_{2-10~{\rm keV}} \sim 2\times10^{33}$~\ergs), with a very low Eddington ratio, L/L$_{\rm Edd} \sim 10^{-8}$ to $10^{-11}$ \citep[Figure \ref{fig:fov}; e.g.,][]{Baganoff01, Baganoff03, Xu06, Genzel10, Wang13}. The size of the X-ray emitting region is similar to the black hole's Bondi radius ($\sim6000$ au, roughly $10^{5}$ times the Schwarzschild radius), through which the accretion rate is $10^{-6}$\msun\ yr$^{-1}$ \citep{Melia92,Quataert02}. The gas reservoir likely originates from captured winds from the S stars or the cluster of massive OB stars surrounding \sgra\ \citep{Cuadra05,Cuadra06,Cuadra08,Yusef-Zadeh16,Russell17}, only a fraction of which reaches the SMBH --- the vast majority of the material is apparently ejected in an outflow \citep[e.g.,][]{Wang13}.

Ambitious X-ray campaigns with \Chandra, \XMM, \Swift, and \Nustar\ have targeted \sgra\ and shown that its emission is relatively quiescent and has been stable over two decades, punctuated by approximately daily X-ray flares \citep{Baganoff01, Goldwurm03, Porquet03, Porquet08, Belanger05, Nowak12, Degenaar13, Neilsen13, Neilsen15, Barriere14, Ponti15, Mossoux16, YuanWang16, Zhang17, Bouffard19}. The quiescent component is well-modeled by bremsstrahlung emission from a hot plasma with temperature $T\sim7 \times 10^7$ K and electron density $n_e$ $\sim$100 cm$^{-3}$ located near the Bondi radius \citep{Quataert02,Wang13}. 

\citet{Baganoff01} identified the first \sgra\ X-ray flare with \Chandra\ and modeled it using a power-law spectrum with $\Gamma = 1.3^{+0.5}_{-0.6}$ and $N_H = 4.6\times10^{22}$ cm$^{-2}$, but pile-up led to known biases in the analysis. X-ray flares were subsequently observed by \XMM\ \citep{Goldwurm03,Porquet03,Porquet08,Belanger05,Ponti15,Mossoux16,Mossoux17} for which pile-up is less severe, and \Swift\ \citep[e.g.,][]{Degenaar13}, establishing X-ray flares as an important emission mechanism. 

These findings motivated the \Chandra\ \sgra\ X-ray Visionary Program\footnote{\sgra\ \Chandra\ XVP: http://www.sgra-star.com/} (XVP) in 2012, which utilized the high energy transmission gratings (HETG) to achieve high spatial and spectral resolution, and to avoid complications from pile-up. The XVP uncovered a very bright flare reported by \citet{Nowak12}, and provided a large, uniform sample of fainter flares \citep{Neilsen13,Neilsen15}. {\it NuSTAR} observations have since confirmed that \sgra's X-ray flares can have counterparts at even higher energies \citep[up to $\sim79$ keV;][]{Barriere14,Zhang17}. 

Detection of large numbers of X-ray flares has allowed us to establish correlations between the flare durations, fluences, and peak luminosities \citep{Neilsen13, Neilsen15}. However, the physical processes driving the flares are still up for debate. Magnetic reconnection is a favored energy injection mechanism and can explain the trends, but current observations cannot rule out tidal disruption of $\sim$kilometer-sized rocky bodies \citep{Cadez08, Kostic09, Zubovas12}, stochastic acceleration, shocks from jets or accretion flow dynamics, or even gravitational lensing of hot spots in the accretion flow \citep[e.g., recent efforts by][]{Dibi14, Dibi16, Ball16, Ball18, Karssen17, GRAVITY18a}. The radiation mechanisms are also debated, though synchrotron radiation is favored, particularly in recent joint X-ray and near-infrared (NIR) studies \citep[e.g.,][]{Ponti17, Zhang17, Boyce18, Fazio18}.

X-ray monitoring of \sgra\ continued post-XVP, targeting the pericenter passage of the enigmatic object ``G2'' \citep[e.g.,][]{Gillessen12,Plewa17,Witzel17}. These surveys captured the first outburst from the magnetar \magnetar, separated from \sgra\ by only 2\secspt4 \citep[Figure \ref{fig:fov};][]{Kennea13, Mori13, Rea13, CotiZelati15, CotiZelati17} --- this bright source complicated X-ray monitoring for all but the {\it Chandra X-ray Observatory}, whose spatial resolution is sufficient to disentangle it from \sgra. 

We present an analysis of two of \sgra's brightest X-ray flares, discovered during the 2013--2014 post-XVP \Chandra\ campaigns. In \S\ref{sec:obssec} we describe the \Chandra\ X-ray observations, in Sections \ref{sec:lcs}, \ref{sec:specmod}, and \ref{sec:psds} we present the X-ray lightcurves, spectral models, and power spectra of these two spectacular events. We discuss our findings in \S\ref{sec:disc} and conclude briefly in \S\ref{sec:concl}.

\section{Observations}
\label{sec:obssec}

We report two extremely bright X-ray flares from Sgr A* discovered during \Chandra\ ObsID 15043 on 2013 September 14 (F1) and ObsID 16218 on 2014 October 20 (F2). The observations and quiescent and flare characteristics are summarized in Tables \ref{tab:gtis} and \ref{tab:spec}, and visualized in Figures \ref{fig:fov} and \ref{fig:lcs}. Observations were acquired using the ACIS-S3 chip in FAINT mode with a 1/8 subarray. The small subarray helps mitigate photon pile-up, and achieves a frame rate of 0.44 s (vs. \Chandra's standard rate of 3.2 s). 

\Chandra\ data reduction and analysis are performed with CIAO v.4.8 tools \citep[CALDB v4.7.2;][]{Fruscione06}. We reprocess the level 2 event files, apply the latest calibrations with the \Chandra\ {\tt repro} script, and extract the 2--8 keV light curves, as well as X-ray spectra, from a circular region with a radius of 1\secspt25 (2.5 pixels; Fig. \ref{fig:fov}) centered on the radio position of Sgr A*: R.A. $= 17:45:40.0409$, decl. $= -29:00:28.118$ \citep[J2000.0;][]{Reid04}. The small extraction region and energy filter help isolate the flare emission from Sgr A* and minimize contamination from diffuse X-ray background emission \citep[e.g.,][]{Nowak12, Neilsen13}. Light curves and spectra from the magnetar \magnetar\ are also extracted from a circular aperture with radius $=$ 1\secspt3 \citep[Figs. \ref{fig:fov} and \ref{fig:lcs}; see also][]{CotiZelati15,CotiZelati17}. 

\section{Lightcurves}
\label{sec:lcs}

We present lightcurves from \sgra\ and the magnetar \magnetar\ during two bright \sgra\ X-ray flares in Figure \ref{fig:lcs}. The top panels show the flares with 300 s binning, with an additional $\sim$20 ks before and after the outbursts --- the full data sets are roughly double this length (Table \ref{tab:spec}) but do not contain additional flares. \sgra's $2-8$~keV emission is shown in black; the blue and green curves represent the $4-8$~keV and $2-4$~keV lightcurves, respectively. Though each \sgra\ observation shows stable quiescent emission punctuated by a long, high contrast flare, the $2-8$~keV lightcurve from the magnetar \magnetar\ (orange curve) does not indicate variability during these epochs. Hence we attribute the flare emission to the SMBH alone. 

To determine the mean and variance of the quiescent count rates in nonflare intervals, we define quiescent good time intervals (GTIs) separated by $\sim 6$~ks from the flares, shown in Table \ref{tab:gtis}. The quiescent GTIs amount to 29.5~ks and 22.4~ks of exposure for ObsID 16218 and 15043, respectively. For flare F2 we also exclude the secondary flare at $\sim56950.63$~MJD. The gray shaded regions in Fig. \ref{fig:lcs} represent unused intervals not included in either the flare or quiescent GTIs. We define the flare start and stop times as those points where the count rate rises $2\sigma$ above the quiescent mean (marked with the first and last cyan points in the top panels of Figure \ref{fig:lcs}). 

\citet[][]{Ponti15} calculated the relationship between the incident $2-8$~keV count rate and the observed count rate for the \Chandra\ ACIS-S detectors and found that pile-up produces detected count rates up to $\sim 90$\% lower than they should be, even in the 1/8 subarray mode. We use their relation to compensate for the effects of pile-up (see \S \ref{sec:pileup}) and report the fluence, mean, and max count rates for the raw and corrected lightcurves in Table \ref{tab:spec}.

\begin{deluxetable}{lcccc}
\tabletypesize{\footnotesize}
\tablewidth{0pt}
\tablecaption{{\it Chandra} \sgra\ Quiescent Properties}
\tablehead{{ObsID}  & {start} & {stop} & {mean} & {variance}\\
    & (MJD) & (MJD) & (cnts/s) & (cnts/s)} 
\startdata
15043   & 56549.22 &  56549.59  & $9.4\times10^{-3}$ & $0.9\times10^-3$\\
\multirow{2}{*}{16218} & 56950.36 & 56950.49 & \multirow{2}{*}{$5.9\times10^{-3}$}  & \multirow{2}{*}{$0.6\times10^{-3}$} \\
  & 56950.67 &  56950.82 & & \\
\enddata
\tablenotetext{}{GTIs for the quiescent intervals in ObsID 15043 and 16218, and the corresponding mean and variance in the quiescent count rate. Flare properties are detailed in Table \ref{tab:spec}.}
\label{tab:gtis}
\end{deluxetable}

\subsection{F1: 2013 September 14}
F1 is the brightest \sgra\ X-ray flare yet detected by \Chandra\ (Figure \ref{fig:lcs}, left panels). It has a double-peaked morphology with a rapid rise and a slight shoulder in the decline. The rise time of $1500$~s is measured from the start of the flare to the maximum of the first peak (first two cyan points; first peak occurs at 56549.10 MJD). The decay time of 2500~s is measured from the maximum of the second peak to the end of the flare (second two cyan points; second peak at 56549.12 MJD). The entire flare lasts 5.7 ks (approximately 2 hr), and the two bright peaks are separated by 1.8 ks ($\sim$30 minutes).

F1 has a mean raw count rate of $0.48\pm0.01$~counts~s$^{-1}$, determined by averaging photons within the flare-only GTI, compared to the mean quiescent count rate of $0.009\pm0.01$~counts~s$^{-1}$. The two individual sub-peaks reach their maxima at $1.04\pm0.06$ and $0.93\pm0.06$~counts~s$^{-1}$, or a factor of $2.2\pm0.1$ and $1.9\pm0.1$ times the average flare-only count rate, respectively. (These are raw count rates, pile-up-corrected count rates are listed in Table \ref{tab:spec} and discussed further in \S\ref{sec:pileup}.)

\subsection{F2: 2014 October 20}
F2 is a factor of $\sim2.5$ less luminous than F1, and yet is the second-brightest \sgra\ \Chandra\ X-ray flare reported (Figure \ref{fig:lcs}, right panels). This flare also has complex morphology, though it does not show the distinctive double-peaked structure seen in F1 (peak occurs at 56950.58 MJD). There is a small shoulder during its rise (which may be a precursor flare or a substructure), after which it smoothly peaks and decays back to quiescence. The flare lasts approximately 3.4~ks ($\sim1$~hr) and its morphology is similar to the bright flare reported by \citet{Nowak12}. The maximum F2 flare raw count rate of $0.52\pm0.04$~counts~s$^{-1}$ is a factor of $2.4\pm0.2$ times its raw mean count rate of $0.22\pm0.01$~counts~s$^{-1}$. The mean quiescent count rate is $0.006\pm0.01$~counts~s$^{-1}$.
Approximately 2~ks after the end of the bright flare we detect another, smaller peak, rising $\sim 4\times$ above the mean quiescent level.

\subsection{Lightcurve Morphologies}
\label{sec:morph}
The lightcurve models and best fit parameters for F1 and F2 are described in detail in Appendix B; we summarize them here. The model for F1 is composed of four symmetric Gaussians: one for each of the two tall peaks, one for the last small subpeak, and one to model the asymmetry in the rise and decay. The model for the F2 lightcurve has two components: a symmetric Gaussian representing the main flare and a skewed Gaussian to model the preflare shoulder. The F2 fit also includes a third skewed Gaussian to model the small secondary flare near $\sim56950.63$~MJD. Both models contain a constant component for the quiescent contribution. Since the hardness ratios are relatively constant (\S \ref{sec:hr}), we fit these $2-8$~keV models to the $2-4$~keV and $4-8$~keV lightcurves as well. The best-fit, composite models are shown as solid red lines in the top panels of Figure \ref{fig:lcs} and in Figure \ref{fig:alcs}.

\subsection{Hardness Ratio}
\label{sec:hr}
\begin{figure*}
\center{
\includegraphics[scale=0.40]{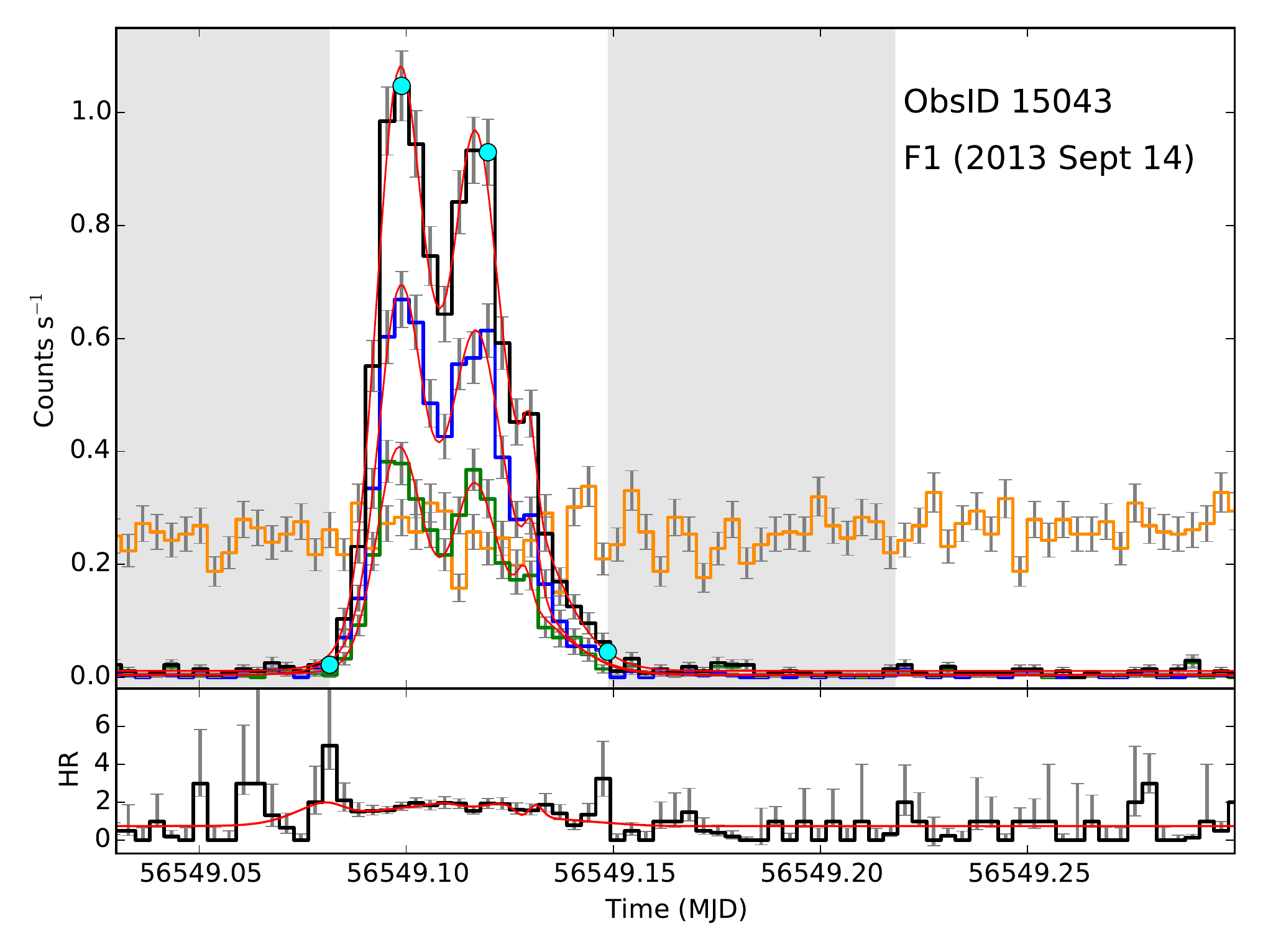}
\includegraphics[scale=0.40]{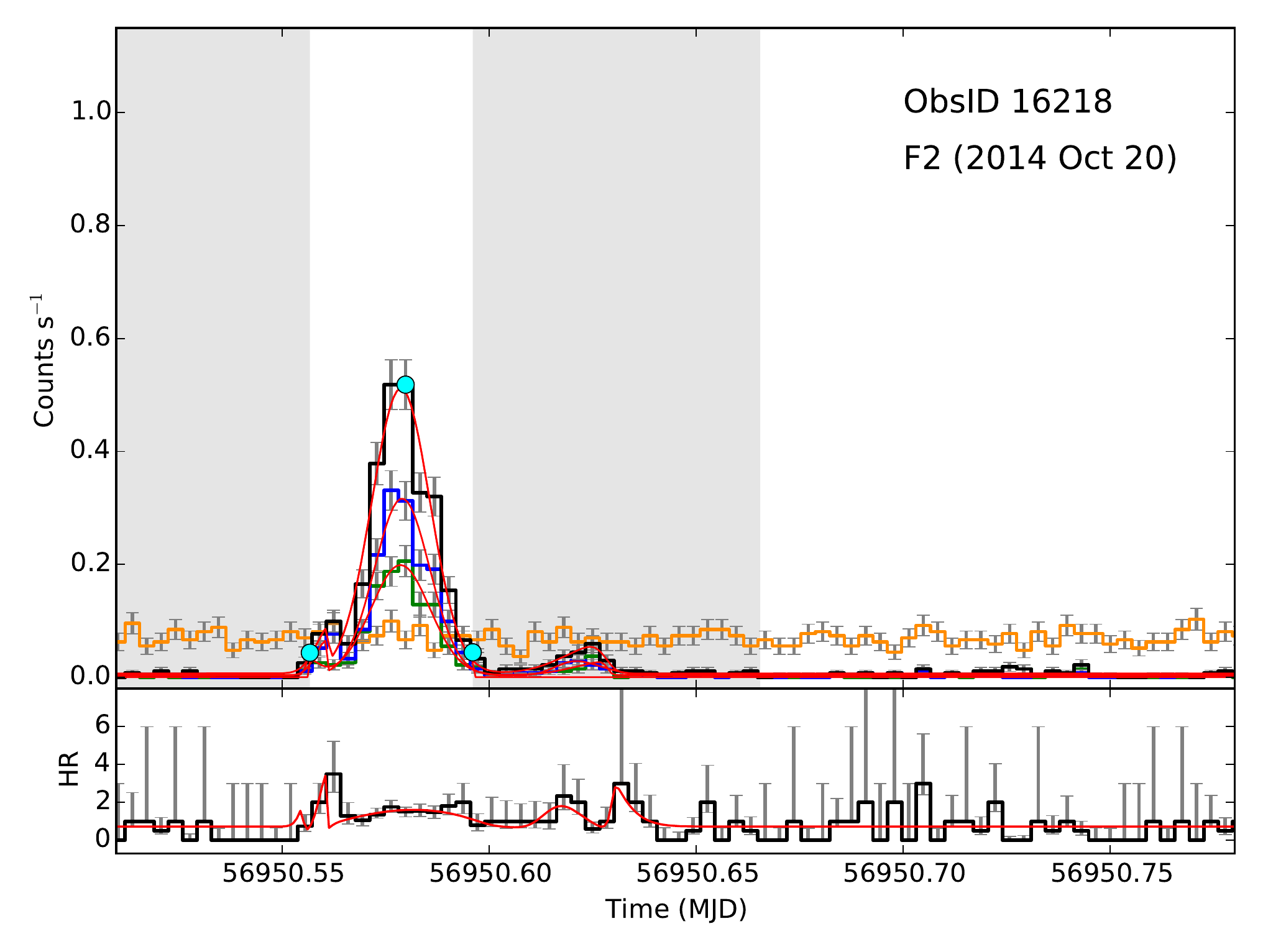}
\includegraphics[trim={0 0.1cm 0 0.5cm},clip,scale=0.3]{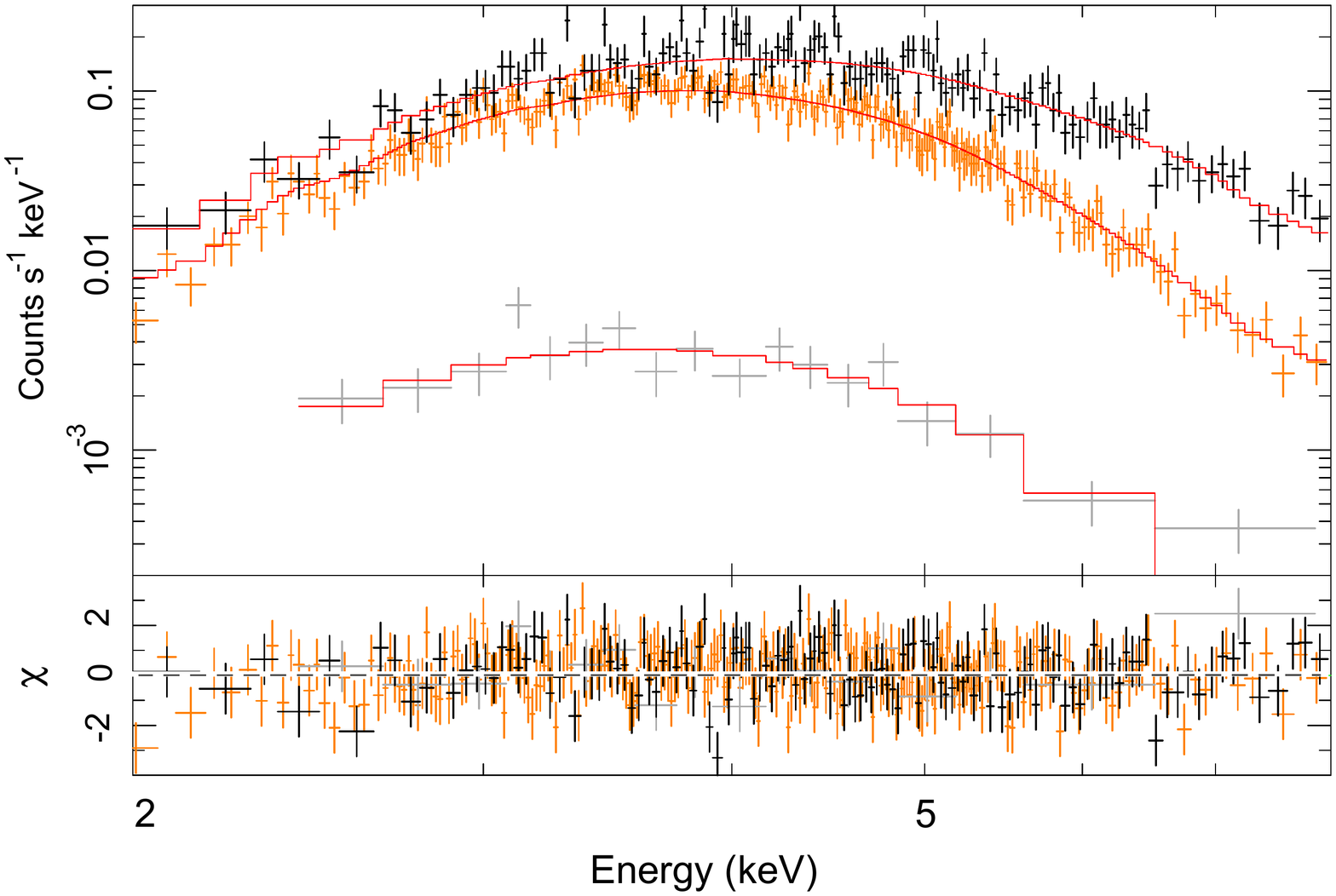}
\includegraphics[trim={0 0.1cm 0 0.5cm},clip,scale=0.3]{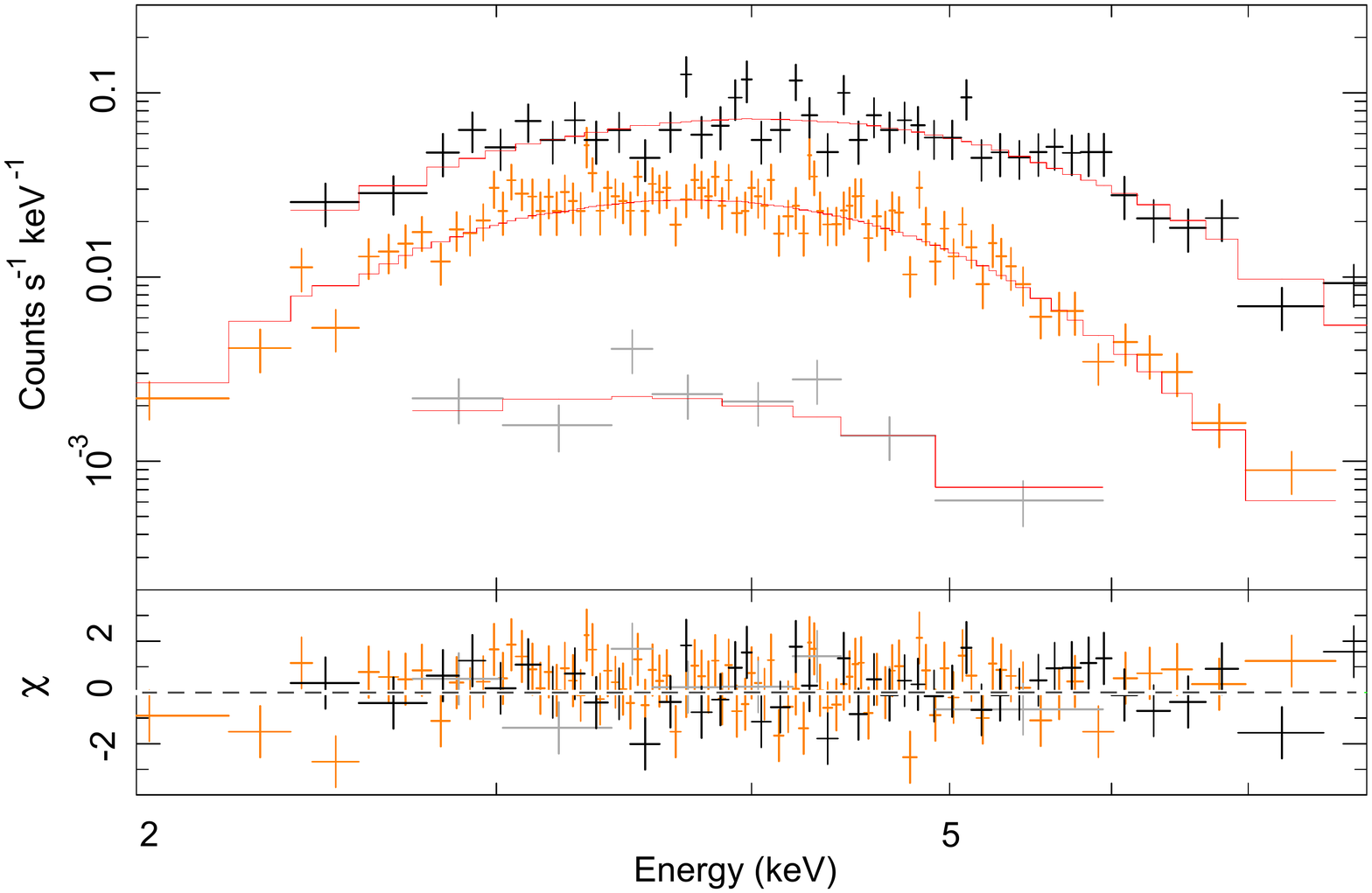} 
\includegraphics[scale=0.40]{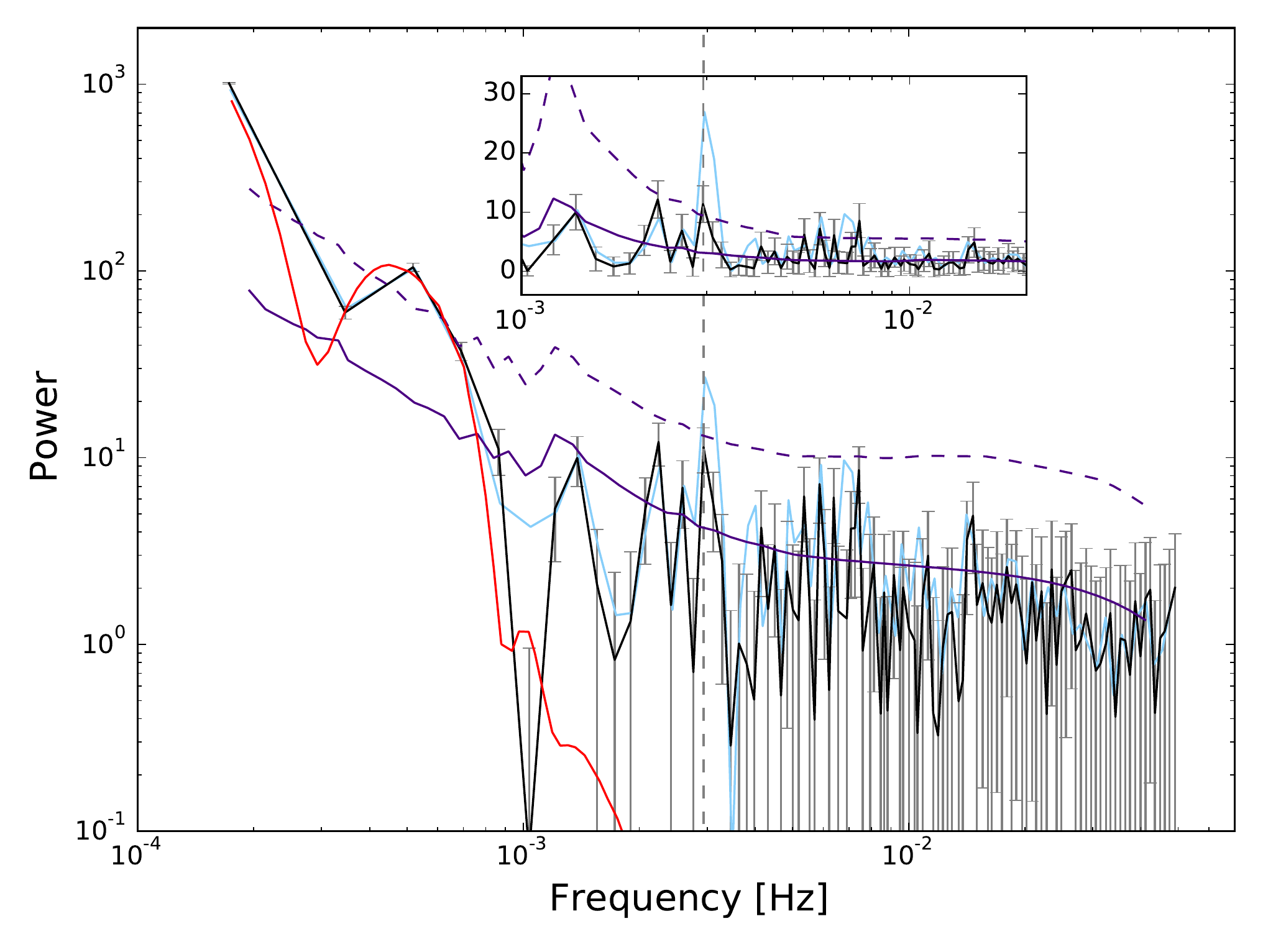}
\includegraphics[scale=0.40]{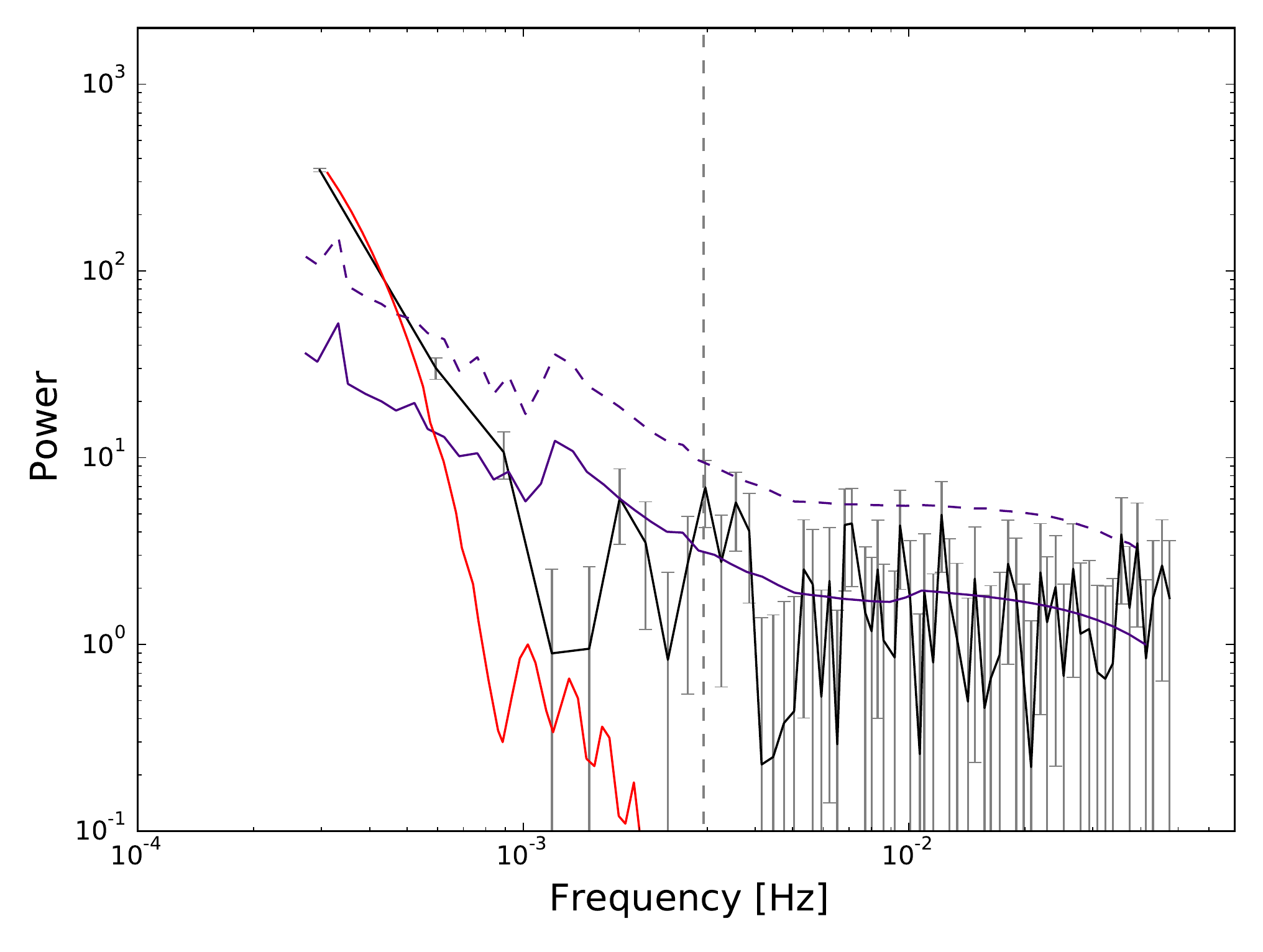}
}
\figcaption{(\textit{\textbf{Top}}) \Chandra\ lightcurves in 300 s bins for the 2013 September 14 flare F1 in ObsID 15043 ({\it left}\,) and the 2014 October 20 flare F2 in ObsID 16218 ({\it right}\,); no pile-up correction has been applied, see instead Fig. \ref{fig:alcs}. The black, blue, and green curves represent emission from \sgra\ in the $2-8$~keV (full), $2-4$~keV (soft), and $4-8$~keV (hard) bands, respectively; the orange line shows the $2-8$~keV emission from the magnetar \magnetar. Smooth red lines are model fits to the \sgra\ lightcurves (\S \ref{sec:morph}, see also Appendix \ref{app:b}). The flare start and stop times and brightest peaks are indicated by small cyan dots. The gray shaded areas mark times that fall outside flare and quiescent GTIs. The lower subpanels show the hardness ratio (HR $=$ hard\,/soft), overplotted with the hard-to-soft model ratio from the lightcurves (solid red lines). 
(\textit{\textbf{Middle}}) Pile-up-corrected X-ray spectra for F1 ({\it left}\,) and F2 ({\it right}\,). The \sgra\ flare component is in black, quiescence in gray, and the magnetar in orange, with spectral models overplotted in red (see Table \ref{tab:spec}). The lower subpanels show the $\chi^2$ residuals for each component fit. 
(\textit{\textbf{Bottom}}) Black curves are the power spectra derived via a Fast Fourier Transform for the two bright X-ray flares. The pale blue curve for F1 shows the PSD including instrumental effects and the dashed gray line marks the spurious peak in F1 at $\nu = 2.99\pm0.14$ mHz; this section of the PSD is also shown in the inset with the y-axis on a linear scale. We mark this frequency in the F2 PSD for reference.
The dark purple solid and dashed lines mark the 50\% ({\it bottom}\,) and 90\% ({\it top}\,) confidence intervals for a combined red noise ($\beta = 1$) plus white noise spectrum (\S\ref{sec:errors1} and Appendix \ref{app:a}) --- the solid red curve shows the FFT of our fit to the $2-8$~keV lightcurve (top panels). }
\label{fig:lcs}
\end{figure*}

We calculate flare hardness ratios (HRs) as the ratio of the numbers of hard ($4-8$~keV) to soft ($2-4$~keV) photons, and plot them in the subpanels of the lightcurves at the top of Figure \ref{fig:lcs}. Errors on the hardness ratios are calculated using the Bayesian analysis described in \citet{Park06} and are shown in the subpanels below the lightcurves in Fig. \ref{fig:lcs}. Bins containing zero counts in the $4-8$~keV lightcurve are replaced with the median quiescent count rate for plotting purposes only. The hardness ratio for F1 exhibits two peaks that coincide with the flare start/stop times. F2 has a similar structure, though it shows a more gradual increase in hardness ratio at the start of the flare, and the feature at the end of the flare is not as prominent. These features in the HR have large errors and we therefore do not consider them significant.

The HR between these two bright flares is fairly uniform, $1.7\pm0.3$ and $1.5\pm0.2$ for F1 and F2, respectively. These are consistent within errors, but somewhat lower than the average flare HR of $\sim2$ for the bright X-ray flare studied by \citet{Nowak12}. Different effective areas for the ACIS-S and HETG instruments may account for this discrepancy. Pile-up also impedes our ability to measure the spectral shape and timing of the flare photons --- we assess the degree of pile-up in our spectral fits in \S \ref{sec:pileup} and \ref{sec:specmod}.

\subsection{Pile-up Correction}
\label{sec:pileup}
Pile-up occurs when two or more photons strike a pixel during a single CCD readout. The two photons are subsequently recorded as a single event with a summed energy. This information loss reduces the incident count rate and falsely hardens the spectrum for bright X-ray events like F1 and F2. Using a 1/8 subarray for these observations lowers the frame rate to $\Delta t\sim0.44$ s ($0.4$ s for the primary exposure, plus $0.04$ s for readout) from a nominal $\Delta t=3.2$ s, which mitigates the effects of pile-up, but does not fully eliminate them.

To correct the lightcurve for pile-up, we use the formula from the {\it Chandra} ABC Guide to Pileup,\footnote{http://asc.harvard.edu/ciao/download/doc/pileup\_abc.ps}
\begin{equation}
f_r = 1 - \frac{ [exp(\alpha\Lambda) - 1] [exp(-\Lambda)]}{\alpha\Lambda}
\label{eqn:pu}
\end{equation}
where $f_r$ is the fraction of the incident count rate lost due to pile-up, $\Lambda$ is the incident rate, and $\alpha$ is the pile-up grade migration parameter. Here the rates are expressed as counts per frame rather than the standard counts per unit time. We use this parameterization to calculate a pile-up-corrected incident count rate for each $\sim$50~s bin in the lightcurves shown in Figure \ref{fig:alcs}.  As expected, higher count rates are substantially more affected by pile-up degradation. 

\section{Spectral Analysis}
\label{sec:specmod}

Since \sgra\ and \magnetar\ are separated by only 2\secspt4, we minimize spectral cross-contamination by selecting small extraction regions and appropriate GTIs. 
We perform spectral fits with the {\texttt{Interactive Spectral Interpretation System}} v1.6.2-32 \citep[{\texttt{ISIS};}][]{HouckDenicola00}. 
The neutral hydrogen column absorption ($N_H$) is modeled with \texttt{TBnew}, and we use the photoionization cross sections of \citet{verner96} and the elemental abundances of \citet{Wilms00}. Interstellar dust can scatter X-ray photons and is a crucial component in X-ray spectral fitting of Galactic center sources \citep{Corrales16, Smith16, Corrales17, Jin17}. We apply the dust scattering model \texttt{fgcdust} \citep{Jin17} developed for AX J1745.6$-$2901 and tested on X-ray sources in the Galactic center region by \citet{Ponti17}. Spectral fits are detailed in the bottom half of Table \ref{tab:spec}.

\begin{deluxetable*}{lccccccccccc}
\tabletypesize{\footnotesize}
\tablewidth{0pt}
\tablecaption{{\it Chandra} \sgra\ Bright Flare Properties}
\tablehead{\multicolumn{12}{c}{TIMING}}
\startdata
{Flare} & {ObsID} & {Tot} & {Flare} & {Flare} & {Flare} & {Rise} & {Decay} & {Fluence} & {Mean Rate} & {Peak1 Rate} & {Peak2 Rate} \\
{} & {} & {Exp} & {Start} & {Stop} & {Duration} & {Time} & {Time} & {raw/corr} & {raw/corr} & {raw/corr} & {raw/corr} \\
{} & {} & {(ks)} & {(MJD)} & {(MJD)} & {(ks)} & {(s)} & {(s)} & {(cnts)} & {(cnts/s)} & {(cnts/s)} & {(cnts/s)} \\
\hhline{------------}
F1  & 15043 & 45.41 &  56549.08 & 56549.15  & 5.73 &  1500 & 2500  & 2769/3115 & 0.48~/~0.54 & 1.04~/~1.17 & 0.93~/~1.06\\
F2  & 16218 & 36.35 &  56950.56 & 56950.60  & 3.36 &  1700 & 1400  &  734/825 & 0.22~/~0.24 & 0.52~/~0.59 & \ldots \\
\hhline{============} 
\noalign{\vskip 2.5mm}  
\multicolumn{12}{c}{SPECTRAL FITS}\\
\noalign{\vskip 1.mm}  
\hhline{------------}
{} & {} & \multicolumn{6}{c}{--------------------------------------------- Flare -------------------------------------------} & \multicolumn{2}{c}{------------ Magnetar ------------} & {} & {} \\
Flare & State & F$^{\text{abs}}_{2-8}$ & F$^{\text{unabs}}_{2-8}$ & F$^{\text{unabs}}_{2-10}$ & L$^{\text{unabs}}_{2-10}$ & $\Gamma$ & HR & F$^{\text{abs}}_{2-8}\times10^{-12}$ & kT & $\alpha$ & $\chi^2/DoF$ \\
{} & {} & \multicolumn{3}{c}{($\times10^{-12}$~erg/cm$^{2}$/s)} & ($\times10^{34}$~erg/s) & {} & {} & (erg/cm$^{2}$/s) & (keV) &  {} & {} \\
\hline
F1 & Mean & $28.5_{-1.6}^{+1.7}$ & $64.8_{-3.1}^{+3.2}$ & $74.4_{-3.5}^{+3.7}$ & $57.0_{-2.7}^{+2.8}$ & $2.06\pm0.14$ & $1.7\pm0.3$ & $6.9\pm0.2$ & $0.79\pm0.02$ & $0.68_{-0.17}^{+0.19}$ & 514.5/423 \\
F1 & Peak1 & $61.3_{-1.6}^{+1.7}$ & $139.5_{-3.1}^{+3.2}$ & $160.1_{-3.5}^{+3.7}$ & $122.6_{-2.7}^{+2.8}$ & \ldots & \ldots & \ldots & \ldots  & \ldots & \ldots  \\
F1 & Peak2 & $54.8_{-1.6}^{+1.7}$ & $124.8_{-3.1}^{+3.2}$ & $143.2_{-3.5}^{+3.7}$ & $109.7_{-2.7}^{+2.8}$ & \ldots & \ldots & \ldots & \ldots  & \ldots & \ldots  \\
F2 & Mean & $10.8\pm0.9$ & $23.3\pm1.8$ & $26.9\pm1.9$ & $20.6\pm1.5$ & $2.03\pm0.27$ & $1.5\pm0.2$ & $1.5\pm0.1$ & $0.75\pm0.04$ & 0.5(f) & 142.8/139 \\
F2 & Peak1 & $25.7\pm0.9$ & $55.5\pm1.8$ & $64.1_{-1.9}^{+2.0}$ & $49.1\pm1.5$ & \ldots & \ldots & \ldots & \ldots  & \ldots & \ldots \\
\enddata
\tablenotetext{}{The top section contains timing parameters for F1 and F2, and the bottom contains the joint spectral fits. 
(\emph{Top}) The $2-8$~keV fluence, mean, and peak rates each have two entries, one for observed (raw) values, the other for the pile-up corrected (corr) values. These contain both flare and quiescent counts since both contribute to pile-up during the flares.  The rise time for flare F1 is measured from the flare start time to the maximum of the first peak, the decay time is measured from the maximum of the second peak to the flare stop time. For the double-peaked flare F1 each of the two peak rates are listed. 
(\emph{Bottom}) \sgra\ flares are fit with the model \texttt{fgcdust$\times$TBnew$\times$(powerlaw$_f$+powerlaw$_q$+blackbody)}, where the two power laws fit the flare and quiescent spectral components during the flare, and the blackbody fits the contribution from the magnetar. The magnetar-only fits employ a single blackbody and utilize data exclusively during \sgra's quiescent periods to avoid contamination. 
Dust scattering and absorption are included in both fits. The quiescent PL index is fixed to $\Gamma=3.0\pm0.2$ \citep{Nowak12, Neilsen13}. 
The flux contribution from the magnetar is estimated to be 3\% and 2.5\% for F1 (ObsID 15043) and F2 (ObsID 16218), respectively. $N\textsubscript{H}$ is fixed to $16.3\times10^{22}$~cm$^{-2}$ and we assume a distance to the Galactic center of 8 kpc. Pile-up ($\alpha$) is a free parameter in the F1 fit, but is fixed to $\alpha=0.5$ for the F2 fit because it could not be constrained by the data (\S\ref{sec:fspec}). 
We list the 2-8 keV absorbed and unabsorbed fluxes, as well as the 2-10 keV unabsorbed flux and luminosity. 
The peak flux values are derived assuming peak/mean flux ratios of $2.2\pm0.1$, $1.9\pm0.1$, and $2.4\pm0.2$ (90\% confidence limits), for F1 peak1, F1 peak2, and the F2 peak, respectively, and errors are combined in quadrature (\S\ref{sec:lcs}). Hardness ratios (HR) are defined as $4-8$ keV/$2-4$ keV (\S\ref{sec:hr}).}
\label{tab:spec}
\end{deluxetable*}

\subsection{\magnetar}
\label{sec:magspec}

We extract the magnetar's spectrum during nonflare intervals to minimize contamination from \sgra. The spectra from ObsID 15043 and 16218 are jointly fit using an absorbed, dust-scattered blackbody model (\texttt{fgcdust$\times$TBnew$\times$blackbody}) with the neutral hydrogen column tied between the data sets. An additional \texttt{pileup} component is added to the model in ObsID 15043 due to the moderately high count rate of the magnetar during this observation. The best-fit absorption column is $N\textsubscript{H} = (16.3\pm0.8) \times10^{22}$~cm$^{-2}$. The blackbody temperatures ($kT$) are allowed to vary between the observations (the magnetar is known to fade and cool on timescales of months to years), yielding values of $kT=0.79\pm0.02$~keV and $kT=0.75\pm0.04$~keV with $2-8$~keV absorbed fluxes of $6.9\pm0.2 \times10^{-12}$~erg~s$^{-1}$~cm$^{-2}$ and $1.5\pm0.1 \times10^{-12}$~erg~s$^{-1}$~cm$^{-2}$, in agreement with \citet{CotiZelati15,CotiZelati17}. These values for the absorption column and the magnetar temperature are adopted in the spectral fits described below. 

\subsection{\sgra\ Quiescence}
\label{sec:qspec}
We extract spectra of \sgra's quiescent emission using the same off-flare GTIs as for \magnetar. We fit the emission with an absorbed power law model that includes dust scattering. Due to the low number of counts ($\sim280$ for ObsID 15043 and $\sim140$ for ObsID 16218), we find that $N_H$ and the power law index ($\Gamma$) are unconstrained when both are left free. Fixing the absorption column to the magnetar value ($N\textsubscript{H} = 16.3\times10^{22}$~cm$^{-2}$) returns a PL index of $\Gamma_q\sim3.7\pm0.5$. This value is still not well constrained, but matches quiescent spectral fits from previous studies, e.g., \citet{Nowak12}. Since \citet{Nowak12} incorporate all available \Chandra\ observations of \sgra\ up to 2012 ($>3$ Ms), which are not complicated by potential magnetar contamination, we adopt their well-constrained PL index $\Gamma_q=3.0\pm0.2$ in all subsequent spectral analyses.
   
\subsection{Sgr A* Flare}
\label{sec:fspec}
We use the on-flare GTIs to extract \sgra\ flare spectra for both observations. The majority of the photons are from the bright \sgra\ flares, but also include counts from \sgra's quiescent emission, plus a small contribution from the nearby magnetar. At the position of \sgra, the magnetar contributes 3\% and 2.5\% of the flux, for F1 and F2 respectively \citep[estimated with the ray-trace simulation {\tt ChaRT/MARX} and approximations of the GC dust scattering halo, e.g.,][and references therein]{Corrales17,Jin17}. We fix the magnetar contribution manually with the \texttt{model\_flux} tool.
We jointly fit these three components with a flare power law (\texttt{powerlaw$_f$}), a quiescent power law (\texttt{powerlaw$_q$}), and a thermal blackbody (\texttt{blackbody}) for the magnetar. The dust scattering model (\texttt{fgcdust}) and absorption (\texttt{TBnew}) are applied to the total spectrum. 
 
Using the dedicated pile-up model in \texttt{ISIS} (see Eqn. \ref{eqn:pu}), we test spectral fits for F1 and F2 with and without pile-up and find that both suffer from pile-up, though we cannot directly constrain the impact for F2 ($\alpha = 0.0_{-0.0}^{+1.0}$) and instead fix it to $\alpha = 0.5$. We fix the absorption column, the quiescent photon index, and the magnetar temperatures to the values from \S\ref{sec:magspec} and \S\ref{sec:qspec}.

These joint spectral fits give flare power-law indices of $\Gamma=2.06\pm0.14$ (F1) and $\Gamma=2.03\pm0.27$ (F2), corresponding to $2-8$~keV absorbed fluxes of $F_{2-8~keV}= 28.5^{+1.7}_{-1.6}\times10^{-12}$~\ergscm2 for F1 and $F_{2-8~keV}= 10.8\pm0.9\times10^{-12}$~\ergscm2 for F2. We show these fits in the middle panels of Fig. \ref{fig:lcs} and report the fit parameters in Table \ref{tab:spec}, together with the derived fluxes and luminosities for the flares.

As a check on our analysis, we subtract the off-flare photons from the same 1\secspt25 \sgra\ region using the quiescent GTI. This removes contributions from the magnetar, \sgra's quiescent emission, as well as contamination from other sources in the crowded region (e.g., Sgr~A~East). We fit these photons with a simple absorbed power-law model \texttt{fgcdust$\times$TBnew$\times$power-law$_f$}, fix the absorption column, and apply a pile-up model as above. The resultant power-law indices are $\Gamma=1.83\pm0.13$ and $\Gamma=1.93\pm0.30$ for F1 and F2, respectively, in good agreement with our previous findings. This is a useful check on our joint spectral fits, but pileup models may not work properly on these subtracted spectra and are likely to result in harder powerlaw indices (\S\ref{sec:pileup}). We thus prefer the joint spectral fits and adopt those values throughout the remainder of the text and in Table \ref{tab:spec}.

\section{Power Spectra}
\label{sec:psds}

We search for short timescale variability by creating power spectral densities (PSDs) for flares F1 and F2. We perform a fast Fourier transform (FFT) with the \citet{Leahy83} normalization on 10~s binned lightcurves from the flare-only GTIs. The PSDs for F1 and F2 are shown in the bottom panels of Figure \ref{fig:lcs}. The power spectra are complex, with broad features at lower frequencies and smaller-scale modulation at higher frequencies.  

\subsection{PSD Noise Analysis}
\label{sec:errors1}

We examine the broad, low-frequency features by transforming the lightcurve models described in \S\ref{sec:morph} (and Appendix \ref{app:b}) using the same FFT procedure. The resultant model PSDs are overplotted in red in the bottom panels of Figure \ref{fig:lcs}. In both ObsIDs the power spectrum from the flare is well matched to the low-frequency broad curves, but with a sharp drop in power at $\sim10^{-3}$~Hz. 

We investigate the high-frequency features at $>10^{-3}$~Hz to determine whether they can be attributed to short period variability, e.g., quasi-periodic oscillations (QPOs), vs. noise in the lightcurve or aliasing, which can also produce sharp artificial features. We follow a method similar to that of \citet{TimmerKoenig95} \citep[see also implementation by][]{Mauerhan05} and generate simulations where there is no signal. For each flare we create 1000 data sets comprised solely of red and white noise, perform an inverse FFT, and scale the resultant lightcurves so that they have the same variance and sampling as the flare lightcurve. Since the Timmer and Koenig algorithm is prone to windowing, we generate simulated lightcurves $\sim$10 times longer than our flare lightcurves. The simulated data sets are then transformed into a collection of power spectra.

Our choice of noise spectra is motivated by the detection of red noise in past observations of \sgra\ \citep[e.g.,][]{Mauerhan05,Meyer2008,Do09,Witzel12,Witzel18,Neilsen15}. Two values are tested for the red noise slope ($\beta=$1.0, 2.0), which span the range of slopes detected in previous observations \citep{Yusef-Zadeh06, Meyer2008} and simulations \citep{Dolence2012}. At higher frequencies the flare PSD curves taper off into flat Poisson noise, and we therefore also include white noise ($P(f) \sim f^0=1$) in our simulations for a total noise model of $P(f)=f^{-\beta}+1$. The simulated PSD curves for two choices of $\beta$ are shown in Appendix \ref{app:a}; for clarity only the results for $\beta=1.0$ are plotted in the bottom panels of Figure \ref{fig:lcs}, where the $50\%$ and $90\%$ confidence intervals are shown as purple solid and dashed lines, respectively. 

F1 has three features that rise above the $90\%$ confidence interval: two broad curves at lower frequencies and a sharp peak at $\nu\sim3$~mHz (pale blue curve, most prominent in the inset in the bottom left panel of Fig. \ref{fig:lcs}). The first two broad features at $\nu\sim0.17$~mHz (5.7~ks) and $\nu=0.49$~mHz (2.0~ks) arise from the large-scale features in the lightcurve captured by our model, but we cannot attribute the third, narrower peak to the gross morphology of the flare, as it is located well after the flare envelope model drops in power. We fit this peak with a Lorentzian centered at $\nu = 2.99\pm0.01$~mHz, with a width of $\delta\nu = 0.26\pm0.04$~mHz, and a quality factor of $q=\nu/\delta\nu=11.3$. We indicate this frequency on the PSDs as a vertical dashed line in the bottom panels of Figure \ref{fig:lcs}, but argue in \S{\ref{sec:badpix}} that it is an instrumental artifact. 
The residual lightcurves and PSDs in Fig. 3 (dark gray lines) show the result of subtracting this large-scale component and indicate the remaining unmodeled structure. Hence, removing the flare envelop (fitted in the time domain) before searching for high frequency features introduces a spurious signal in the PSD and we do not pursue this approach further here. The only portion of the F2 PSD that rises above the 90\% confidence interval is the flare envelope at low frequencies, there are no significant high-frequency features. We mark the $2.99$~mHz position in the F2 PSD for comparison only.

\subsection{Bad Pixel Analysis}
\label{sec:badpix}
The \Chandra\ spacecraft dithers its pointing position to reduce the effects of bad pixels, nodal boundaries, and chip gaps during an observation, as well as to average the quantum efficiency of the detector pixels\footnote{http://cxc.harvard.edu/proposer/POG/html/chap5.html\#tb:dither}. This observational requirement can introduce spurious timing signatures at either the pitch (X) or yaw (Y) dither periods or their harmonics.

For ObsID~15043 we create an image binned in chip coordinates and extract photons from the same 1\secspt25 region centered on \sgra\ from the level 1 and level 2 event files. Both show the standard Lissajous dither pattern, but also two pixel columns that have been flagged as bad and thus excised in the standard level 2 files. As the \sgra\ extraction region crosses this sharp nodal boundary, the masked pixels create a discontinuity in the lightcurve that manifests as a peak in the power spectrum of F1, at half the period of the yaw dither (because \sgra\ crosses the column of bad pixels twice per oscillation). The \sgra\ extraction region in ObsID 16218 does not overlap a bad column, and \magnetar\ is similarly unaffected in both ObsIDs. Analyses of the \sgra\ and \magnetar\ lightcurves extracted from ObsID 16218 do not reveal any flagged pixels.

In light of this finding, we create new lightcurves for F1 in ObsID 15043 without removing the bad pixel columns, and use these for all subsequent timing analysis. We adopt the extraction region, lightcurve binning, and GTI filtering described in \S\ref{sec:lcs}. The resultant power spectrum is shown in black in the bottom left panel of Figure \ref{fig:lcs}, while the pale blue curve represents the standard level 2 \Chandra\ filtering with the flagged columns removed. The peak at $\sim3$ mHz is no longer detected and we conclude that the possible QPO in F1 arises from instrumentation effects and cannot be attributed to \sgra.

\begin{figure*}
\center{
\includegraphics[scale=0.62,trim={0 0 1cm 1cm},clip]{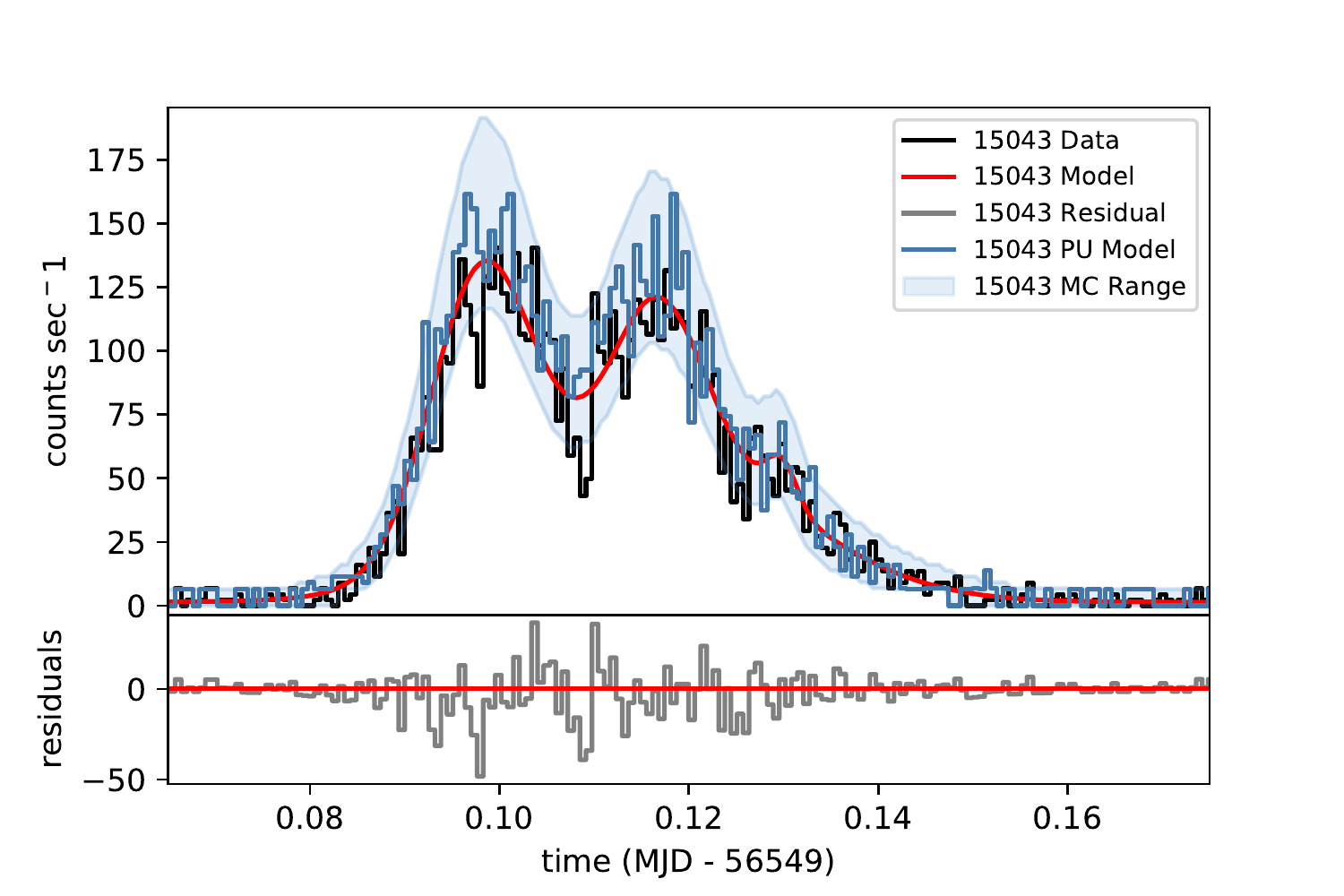}
\includegraphics[width=245pt,trim={0 0 0.5cm 0.5cm},clip]{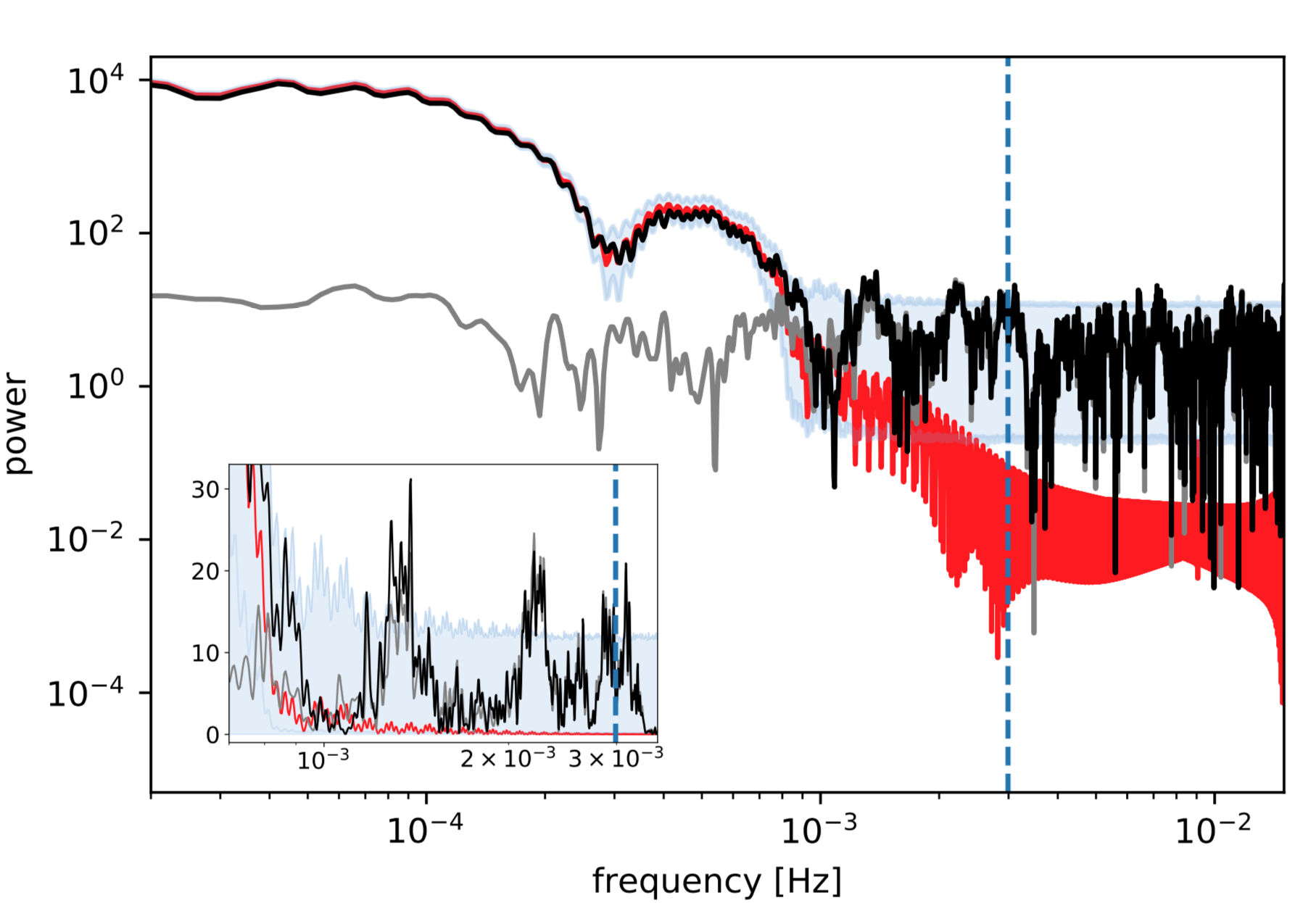}
\includegraphics[scale=0.62,trim={0 0 1cm 1cm},clip]{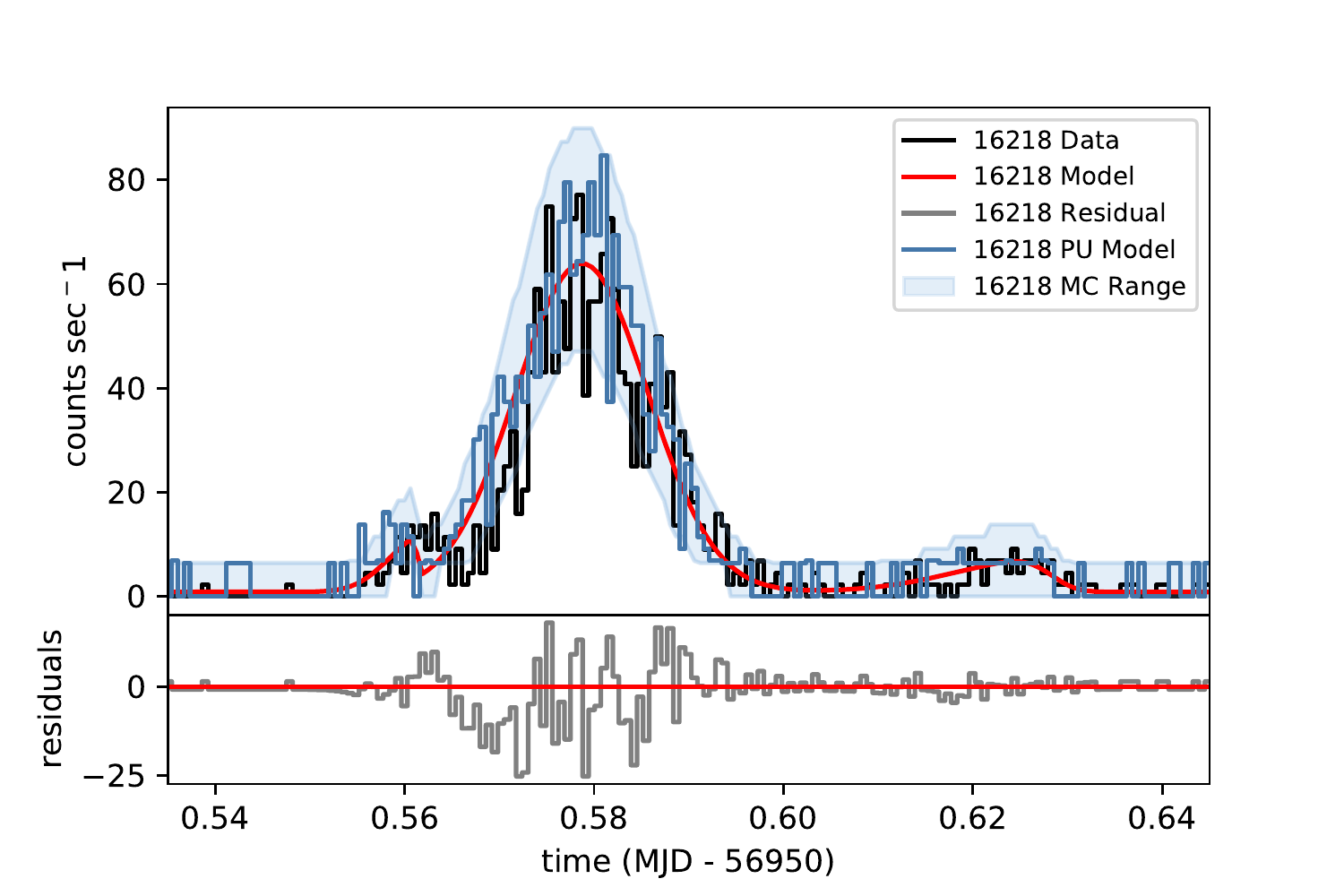}
\includegraphics[width=245pt,trim={0 0 0.5cm 0.5cm},clip]{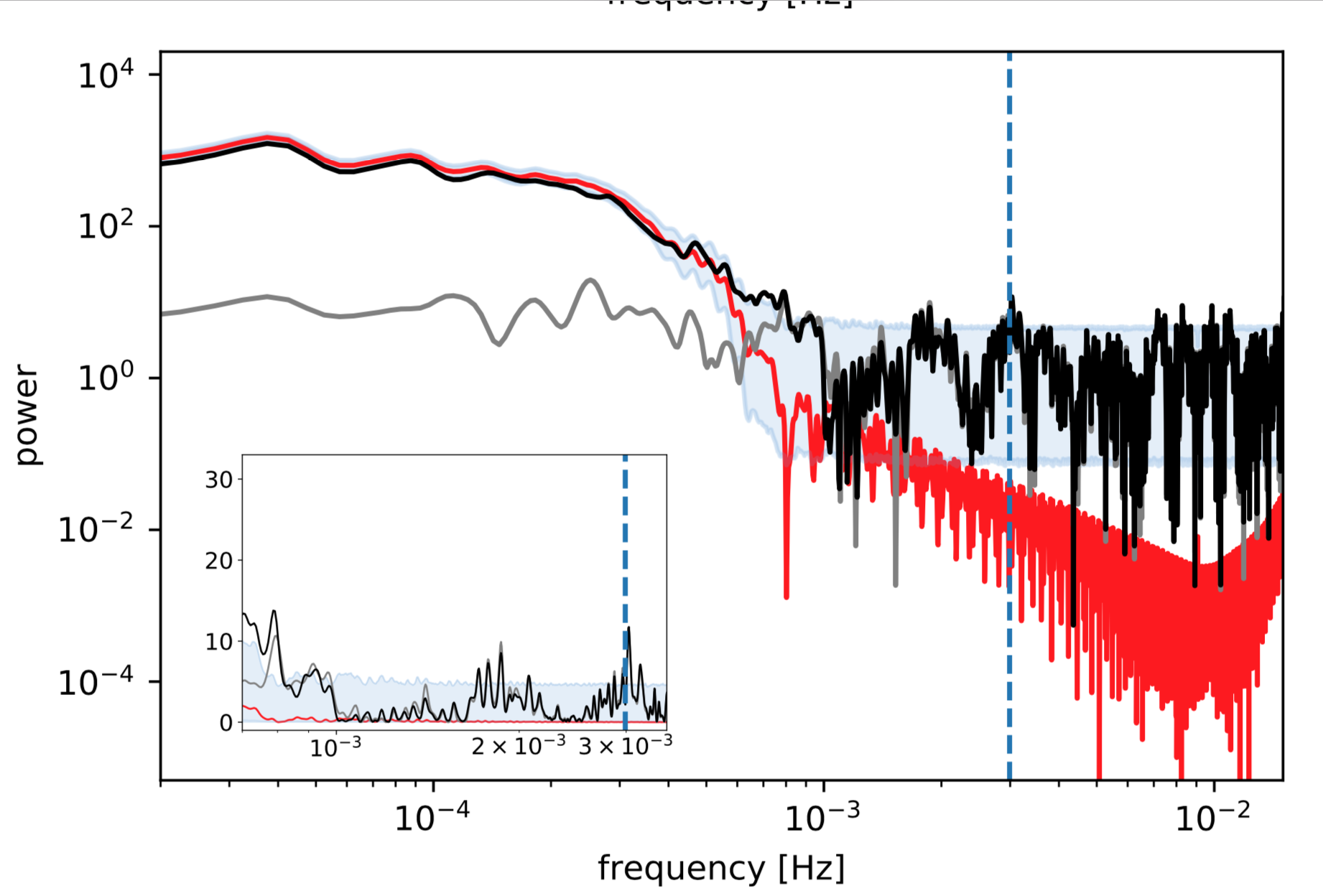}
}
\figcaption{(\textit{\textbf{Left Panels}}) The F1 and F2 lightcurves (black line) and the best-fit model (red curve), binned to $\sim$50 s (vs. the 300 s binning shown in Fig. \ref{fig:lcs}). The residual lightcurve (data $-$ model) is plotted in the lower subpanel (gray line). The pile-up-corrected models are plotted as dark blue lines. Overplotted are the 5--95\% error intervals from 10,000 Monte Carlo simulations of the pile-up-corrected model, generated by adopting a randomly generated Poisson error with a value equal to the larger of either the number of counts in that frame or an expectation value based on the average quiescent count rate (\S \ref{sec:errors2}; light blue shaded regions); binning is performed after the lightcurve simulation is complete.  
(\textit{\textbf{Right Panels}}) The L--S periodograms resulting from the measured and simulated lightcurves (line colors as in left-hand panel). 
} 
\label{fig:alcs}  
\end{figure*}

\subsection{PSD Permutation Error Analysis}
\label{sec:errors2} 
To investigate the high frequency features still apparent in the PSDs, we perform a Monte Carlo (MC) permutation to simulate the errors on the X-ray lightcurves and PSDs. We create 10,000 MC simulations of the full, unbinned \Chandra\ lightcurve (rather than the flare-only intervals) by drawing a simulated count rate for each \Chandra\ frame (frame time $= 0.44$ s for a 1/8 subarray), using our model for the counts from \S\ref{sec:morph} and Appendix \ref{app:b} and a randomly generated Poisson error with an expectation value equal to the larger of (1) the number of counts in that frame or (2) a constant expectation value (the expected number of counts) equal to the average quiescent count rate times the frame time, i.e., 0.004 and 0.003 for ObsID 15043 and 16218. This expectation value is critical for frames where zero counts are detected.

We generate a Lomb--Scargle \citep[L--S;][]{Scargle82} periodogram for each simulated model lightcurve, which is subsequently binned (125 frames are combined to create $\sim$50 s bins) and corrected for pile-up. The left upper panels of Figure \ref{fig:alcs} show the measured lightcurves (black lines), unbinned model fits (smooth red lines), and binned, pile-up-corrected models (dark blue lines), as well as the $5-95\%$ confidence intervals from the MC simulations (light blue shaded regions). The lower subpanels on the left show the residual between the data and the model (gray lines). The right-hand panels show the resultant LS periodogram on a log scale for each of these components and the insets highlight a subset of the higher frequencies. The black curves from the measured LS periodograms have small peaks that rise just above the $95\%$ c.l., but these are weak and most likely attributable to instrumental effects, noise, harmonics of the power in the stronger flare envelopes, and/or aliasing (see Appendix \ref{app:a}). 

\section{Discussion}
\label{sec:disc}

Our analysis reveals two of the brightest X-ray flares seen so far from \sgra, with peak $2-10$ keV luminosities of $12.3\times10^{35}$ \ergs\ (F1) and $49.1\times10^{34}$ \ergs\ (F2), durations of 5.7 and 3.4 ks, and total energies $(0.7-3.3)\times10^{39}$ erg. These are as large or larger than the peak luminosities, durations, and energies found for previously detected bright X-ray flares from \Chandra\ \citep[$L_{2-10~{\rm keV}} = 48\times10^{34}$ \ergs, 5.6 ks, $E_{2-10~{\rm keV}} \sim 10^{39}$ erg;][]{Nowak12} and {\it XMM-Newton} \citep[$L_{2-10~{\rm keV}} \sim (12-19)\times10^{34}$ \ergs, $(2.8-2.9)$ ks, $E_{2-10~{\rm keV}} = (0.3-0.5)\times10^{39}$ erg;][]{Porquet08,Nowak12}.

\subsection{Flare Emission Scenarios}
\label{sec:disEmission}

The X-ray spectral indices for F1 and F2 are consistent with $\Gamma \sim 2$, similar to other bright X-ray flares but significantly harder than the spectrum in quiescence \citep[$\Gamma \sim 3$;][]{Nowak12}. Together with constraints from the NIR, the spectrum seems to favor a synchrotron emission scenario wherein the NIR and X-ray spectral indices are set by the particle acceleration process \citep{Markoff01,Dodds-Eden09}. When the NIR is bright, the spectral index is well described by $\nu f_{\nu} \propto \nu^{0.5}$ \citep[e.g.,][]{Hornstein07, Ponti17} and synchrotron models predict a cooling break, such that $\Gamma \sim 2$ at X-ray wavelengths. Evidence of a similar spectral index for bright and faint flares \citep[e.g.,][]{Neilsen13} also supports this interpretation. 

Alternative models invoke Compton processes \citep[e.g.,][]{Markoff01, Eckart08, Yusef-Zadeh12a}. However, because the Compton bump can shift due to changing scattering optical depths or electron energies (or other factors) flare spectra should show a range of spectral indices. Hence there is no clear expectation that a constant NIR spectral index will lead to a constant X-ray photon index. On the other hand, the large difference between NIR and X-ray cumulative distribution functions \citep{Witzel12,Witzel18,Neilsen15} is not easily understood within the synchrotron scenario \citep{Dibi16}. We must continue to build up the statistics of simultaneous IR and X-ray flares to distinguish between these models.

In addition to their spectral properties, X-ray studies of \sgra\ point to a characteristic timescale for bright flares of $\sim 3-6$ ks \citep[detectable at all durations, e.g.,][]{Porquet03,Porquet08,Neilsen13}. This timescale is also consistent with F1 and F2, reported here. If we associate this timescale with a Keplerian orbital period then the orbital radius is $0.3-0.5$ au, roughly two times the radius of the innermost stable circular orbit (ISCO) for a stationary black hole with \sgra's mass and distance \citep{Boehle16,GRAVITY18b}. This is equivalent to $\sim 5 \times$ the Schwarzschild radius \citep[$R_S$; see detailed models in][]{Markoff01, Liu02, Liu04, Dodds-Eden11}. In this scenario, the flare emission is likely to arise from structures very near or at the event horizon.

Alternatively, we can estimate the size of the emission region (the emitting ``spot'') by comparing the total X-ray energy radiated by the brightest flare, $E_{2-10~keV} = 3.3\times10^{39}$~{\rm erg}, to the energy available in the accretion flow. The volume of the accretion flow that contains an equal amount of magnetic energy is approximately $E_B = (B^2/8\pi) (4\pi R_{spot}^3)$. Setting this equal to the flare's total energy radiated in X-rays  ($E_{2-10~keV}$) and assuming $B = B_0~(R_{f}/3 R_S)^{-1}$, where $B_0 = 30$ G is the magnetic field strength scaled to a typical value close to the black hole \citep{Yuan03,Dodds-Eden09,Moscibrodzka09,Dexter10,Ponti17}, and $R_f$ is the flare emission radius, gives 
\begin{equation}
R_{spot} \sim 2 ~\left(\frac{R_f}{3 R_S}\right)^{2/3} \left(\frac{B_0}{30 ~G}\right)^{-2/3} \left(\frac{E_{2-10~keV}}{3.3\times10^{39} ~{\rm erg}}\right)^{1/3} R_S ~.
\end{equation}

Powering this flare requires a relatively large emitting volume, even assuming a $100\%$ efficiency of converting magnetic energy into X-rays. Increasing the flare emission radius makes the relative size of the hot spot smaller, but only slowly. This poses challenges for explaining the rapid changes in the luminosity \citep[e.g., this work,][among others]{Barriere14}, though an increase in the strength of the magnetic field in the flare region could allow a smaller volume. In general, this represents a lower limit on the size of the flare's total emitting region, because it only takes X-rays into account. It is also consistent with a scenario in which the size of the emitting region is larger for brighter flares.

Recent results from the GRAVITY experiment on the Very Large Telescope Interferometer \citep[VLTI,][]{GRAVITY18b}, argue that \sgra's infrared variability can be traced to orbital motions of gas clouds with a period of $\sim$45 minutes (2.7 ks). Using ray-tracing simulations, they match this timescale to emission from a rotating, synchrotron-emitting ``hot spot'' in a low inclination orbit at $\sim 3-5 ~R_S$ \citep{GRAVITY18a}. 

Since near-IR and X-ray outbursts from \sgra\ appear to be nearly simultaneous and highly correlated \citep[e.g.,][]{Boyce18}, these well-matched timescales give further support to hot spot models and to emission scenarios where the NIR and X-ray emission originates near the SMBH's event horizon. 
We caution, however, that this timescale is not unique, and can also be reproduced by the tidal disruption of asteroid-size objects \citep{Cadez08, Kostic09, Zubovas12} and the Alfv{\`e}n crossing time for magnetic loops near the black hole \citep{Yuan03, Yuan04}.

The GRAVITY NIR flare detected on 2018 July 28 is also double-peaked \citep[see Figure 2 of][]{GRAVITY18a}, much like F1 reported in this work. The durations of these two outbursts, $\sim$115 minutes for the NIR and $\sim$97 minutes for the X-ray, are similar and the time between the two peaks, $\sim$40 minutes (NIR) and $\sim$30 minutes (X-ray), are also similar. The ratio of the total flare duration to the time between peaks is approximately a factor of 3 in both cases. Double-peaked flares like these have previously been detected from Sgr A* \citep[e.g.,][]{Mossoux15,Karssen17,Boyce18,Fazio18}, but 
even when they do not have multiple peaks, bright X-ray flares often display asymmetric lightcurve profiles \citep[e.g., this work,][]{Porquet03,Porquet08,Nowak12,Mossoux16}. Further investigation of their statistical and timing properties is ripe for further study.

\begin{figure}
\center{
\includegraphics[scale=0.45]{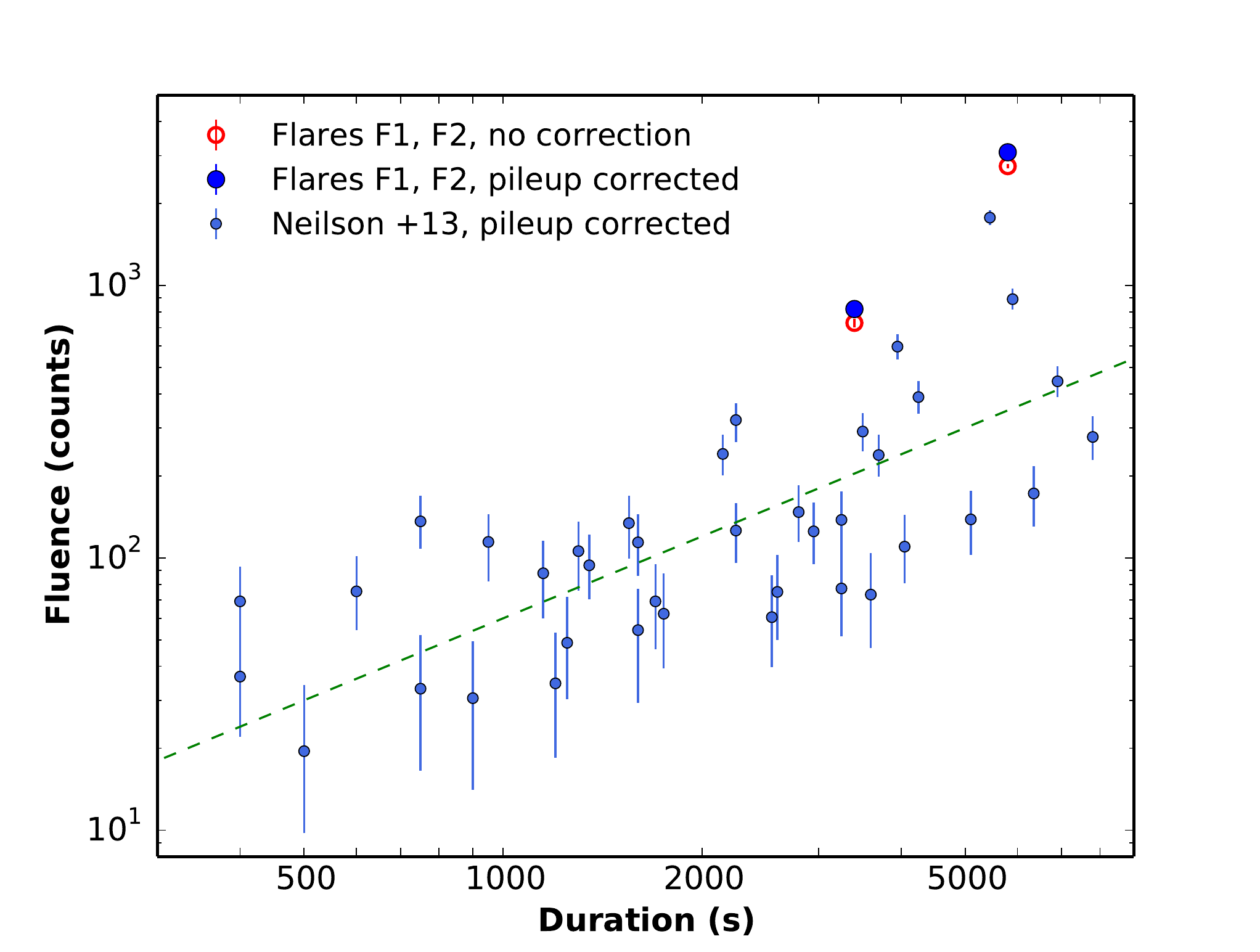}
}
\figcaption{The 2-8~keV flare fluence as a function of flare duration.  The large blue and red data points represent the incident and pile-up corrected data, respectively, for F1 and F2. Smaller blue points represent the flares reported in \citet{Neilsen13}.  Originally reported as fluxes from the HETG configuration, the values were converted to ACIS-S counts for a uniform comparison with the data presented in this work. The dashed green line, from \citet{Neilsen13}, represents flares with luminosities $L=5.4\times10^{34}$~ergs~s$^{-1}$, the approximate demarcation between `bright' flares and `dim' flares.
} 
\label{fig:fluenceRate}
\end{figure}

Despite differences in their morphologies, other flare properties follow more predictable trends. \citet{Neilsen13} present relationships between the fluence, count rate, and duration of $\sim40$ flares from the XVP. To compare F1 and F2 to the XVP flare population, we adjust for the different instrument configurations using PIMMS\footnote{\Chandra\ proposal tools: http://cxc.harvard.edu/toolkit/pimms.jsp.} and scale the reported XVP flare fluences to their equivalent ACIS-S values (Figure \ref{fig:fluenceRate}). F1 and F2 (filled dark blue circles) are considerably brighter than most of the XVP sample, but show the same trend toward higher fluences at longer durations. This again supports a scenario in which the emission region is larger for brighter flares. However, our observations and detection techniques are biased against long-duration, low fluence (i.e., low contrast) flares. Comprehensive flare simulations are under development \citep[e.g.,][]{Bouffard19} and will aide in diagnosing these and other observational biases. 

The flare X-ray hardness ratios and spectral indices appear to be constant, indicating that the spectral slope does not change significantly over the duration of the flare. 
Strengthening this argument, the three brightest X-ray flares detected before our discoveries have photon spectral indices of $\Gamma=2.0^{+0.7}_{-0.6},~2.3\pm0.3,$ and $2.4^{+0.4}_{-0.3}$ \citep{Porquet03,Porquet08,Nowak12}, which are similar to our results of $\Gamma = 2.06\pm0.14$ and $2.03\pm0.27$ for F1 and F2. Recent results from {\it NuSTAR} at higher energies (3-79~keV) provide additional support for a constant PL slope --- \citet{Zhang17} report photon indices of $\Gamma\sim 2.0-2.8\pm0.5$ (90\% confidence) for seven \sgra\ flares. Hence, it appears that \sgra's bright X-ray flares belong to the same population (\S\ref{sec:hr}), and thus are likely to arise from the same energy injection mechanism. 


\subsection{Multiwavelength Coverage}
\label{sec:disMW}

Multiwavelength campaigns targeting \sgra\ often emphasize variability in the X-ray and infrared \citep[e.g., recent studies by][and references therein]{Boyce18,Fazio18}, where emission peaks are nearly simultaneous. Unfortunately, neither F1 nor F2 have near-IR coverage. The GRAVITY Collaboration has demonstrated the power of NIR interferometry in resolving structures only $\sim$ 10 times the gravitational radius of \sgra\ and future simultaneous observations with this instrument and \Chandra\ could offer a powerful probe of how bright X-ray flares are linked to changes in the accretion dynamics near the event horizon.

Longer wavelength observations have hinted at a lag between X-ray flares and their candidate submillimeter or radio counterparts, with the long wavelength data lagging the X-ray. \citet{Capellupo17} report nine contemporaneous X-ray and Jansky Very Large Array (JVLA) radio observations of Sgr A*, including coverage of the F1 flare. The JVLA lightcurve overlapping F1 lasts $\sim 7$ hr and shows significant radio variability at 8-10 GHz ($X$-band) in the form of a 15\% (0.15 Jy) increase during the second half of the observation. The radio lightcurve is not long enough to show a clear start or stop time for the rise, but \citet{Capellupo17} determine a lower limit on the radio flare duration of $\sim$176 minutes, compared to $\sim$100 minutes for the full X-ray flare, with the radio rising before and peaking after the X-ray. This delay could be understood in the context of the adiabatic expansion of an initially optically thick blob of synchrotron-emitting relativistic particles \citep[e.g.,][]{Yusef-Zadeh06}, offering an alternative to the hot spot models discussed above. However, due to the incomplete temporal coverage for F1 and inconclusive correlations between the other radio and X-ray peaks in the study, \citet{Capellupo17} suggest that stronger X-ray flares may lead to longer time lags in the radio, but do not find strong statistical evidence that the radio and X-ray are correlated. 

Observations at 1~mm were also performed at the Submillimeter Array (SMA) a few hours in advance of the F1 flare and on subsequent days as part of a long-term monitoring program \citep{Bower15}. These observations show flux densities that are significantly above the historical average on the date of the F1 flare and for approximately one week afterwards. It is difficult to associate these high flux densities directly with the F1 flare, however, given the infrequent sampling of the SMA light curves and the characteristic time scales of millimeter/submillimeter variability \citep[approximately 8 hr,][]{Dexter14}.

The lack of infrared and patchwork of longer wavelength observations for F1 (and the paucity of multiwavelength data for F2), underscores the importance of continued multiwavelength monitoring of \sgra. 

Toward this end, the {\it Event Horizon Telescope} \citep[EHT;][]{Doeleman08} has recently undertaken its first full-array observations of \sgra\ and M87*. These have offered exquisite, high-precision submillimeter imaging of M87*'s event horizon \citep{EHT2019I} and are poised to do the same for \sgra. Coordinated EHT and multiwavelength observations, including with approved \Chandra\ programs, have begun to inform models for M87* \citep{EHT2019V} and will soon enable a detailed comparison between \sgra's X-ray and submillimeter emission, offering a new understanding of its flares and underlying accretion flow.

\section{Conclusion}
\label{sec:concl}

We have performed an extensive study of the two brightest \Chandra\ X-ray flares yet detected from \sgra, discovered on 2013 September 14 (F1) and 2014 October 20 (F2). After correcting for contamination from the nearby magnetar, \magnetar, and observational and instrumental effects from the Galactic center X-ray background, pile-up, and telescope dither, we report the following key findings:

\begin{enumerate}
\itemsep0em 
\item The brightest flare F1 has a distinctive double-peaked morphology, with individual peak X-ray luminosities more than 600$\times$ and 550$\times$ \sgra's quiescent luminosity. The second bright flare F2 also has an asymmetric morphology and rises 245$\times$ above quiescence.
\item These bright flares have long durations, 5.7 ks (F1) and 3.4 ks (F2), and rapid rise and decay times: F1 rises in 1500 s and decays in 2500 s, while F2 rises in 1700 s and decays in 1400 s. These timescales may correspond to emission from a hot spot orbiting the SMBH, in agreement with new NIR results from the GRAVITY experiment. Their timing properties also imply that the size of the emitting region may scale with the brightness of the flare. 
\item Both flares have X-ray spectral indices and hardness ratios significantly harder than during quiescence, $\Gamma = 2.06 \pm 0.14$ and $\Gamma = 2.03 \pm 0.27$ vs. $\Gamma = 3.0 \pm 0.2$ in quiescence, consistent with previously detected \sgra\ flares and with a synchrotron emission mechanism, though Compton emission models are not ruled out.
\item  Statistical tests of the flare time series, including Monte Carlo simulations of their noise properties, offer little evidence for high-frequency power (e.g., QPOs) in either flare.
\item F1 and F2 are the brightest X-ray flares ever detected by \Chandra, and are among the brightest X-ray flares from \sgra\ detected by any observatory. They appear to deviate in the fluence-duration plane from the distribution of fainter flares detected in the 2011 \sgra\ \Chandra\ XVP. However, low fluence, long duration flares may not exist or may suffer from selection bias, making this finding tentative at present.
\item Neither F1 nor F2 have near-IR coverage, but some (near-)simultaneous coverage of F1 was collected at radio and submillimeter wavelengths. These data point to higher than average flux levels for \sgra\ near the F1 outburst, but do not allow definitive correlation between \sgra's X-ray and longer wavelength behavior. Coordinated multiwavelength campaigns thus continue to be essential for a complete, statistically robust examination of \sgra's multiwavelength variability.
\end{enumerate}

\acknowledgments
The authors thank Charles Gammie, Tiffany Kha, Matthew Stubbs, and Nicolas Cowan for discussions related to this work. We acknowledge the \Chandra\ scheduling, data processing, and archive teams for making these observations possible. This work was supported by \Chandra\ Award Numbers GO3-14121A and GO4-15091C, issued by the {\it Chandra X-ray Observatory Center}, which is operated by the Smithsonian Astrophysical Observatory for and on behalf of the National Aeronautics Space Administration (NASA) under contract NAS8-03060. 

D.H., M.N., and C.O.H. acknowledge support from the Natural Sciences and Engineering Research Council of Canada (NSERC) Discovery Grant. D.H. and M.N. also thank the Fonds de recherche du Qu\'{e}bec--Nature et Technologies (FRQNT) Nouveaux Chercheurs program. D.H. acknowledges support from the Canadian Institute for Advanced Research (CIFAR). M.N. acknowledges funding from the McGill Trottier Chair in Astrophysics and Cosmology. G.P. acknowledges financial support from the Bundesministerium f\"{u}r Wirtschaft und Technologie/Deutsches Zentrum f\"{u}r Luft-und Raumfahrt (BMWI/DLR, FKZ 50 OR 1812, OR 1715 and OR 1604) and the Max Planck Society. S.M. is funded by a VICI grant (639.043.512) from the Netherlands Organisation for Scientific Research.

\facility{CXO}
\software{Python, CIAO}

\appendix
\renewcommand\thefigure{\thesection.\arabic{figure}}    
%
\section{Power Spectra Comparisons}
\setcounter{figure}{0}  
\label{app:a}

\begin{figure}
\center{
\includegraphics[scale=0.42]{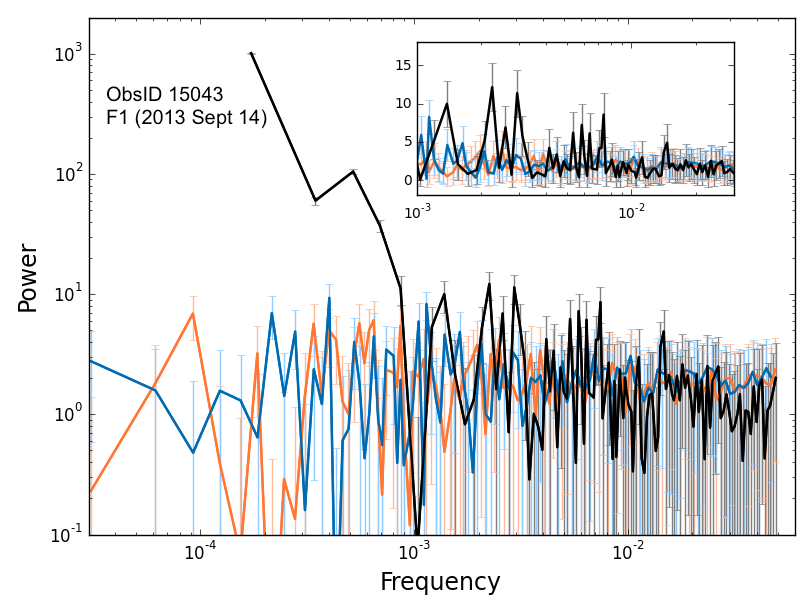}
\includegraphics[scale=0.42]{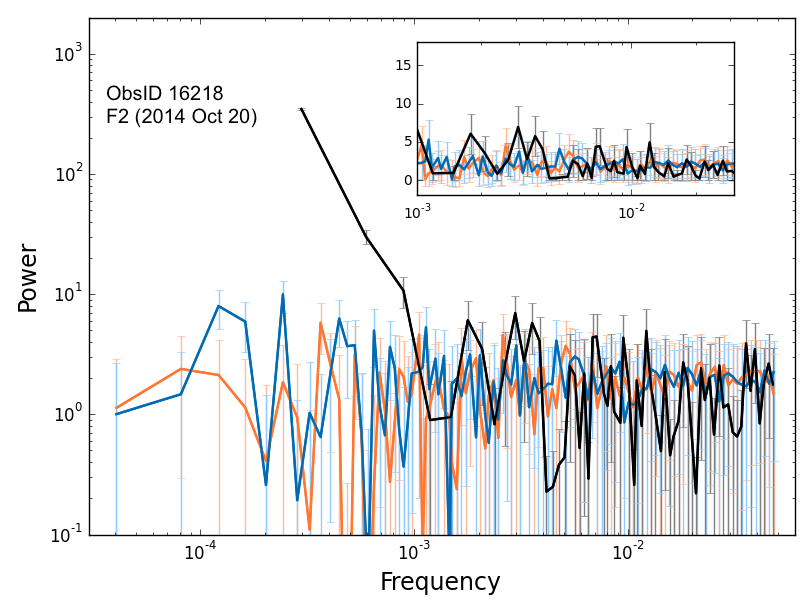}
}
\figcaption{Leahy-normalized power spectra for \sgra\ flares F1 and F2 (black), \sgra\ during quiescence (blue), and the magnetar, \magnetar\ (orange). PSDs are shown for ObsID 15043 (\textbf{\it left}\,) and 16218 (\textbf{\it right}\,).}
\label{fig:appsd_1}
\end{figure}

In addition to the power spectral analysis of the two bright \sgra\ flares presented in Section \ref{sec:psds},  we further investigate \sgra\ during quiescence to compare the respective behaviors at high frequencies. We obtain the quiescent PSD from the GTIs listed in Table \ref{tab:gtis} and process it with the same method as for the flare (see \S\ref{sec:psds}).  For an additional comparison, we also obtain the power spectrum for the magnetar \magnetar\ during the intervals when \sgra\ is in quiescence to minimize contamination.  The resultant Leahy-normalized PSDs are shown in Figure \ref{fig:appsd_1} for ObsID 15043 ({\it left}) and ObsID 16218 ({\it right}). The \sgra\ flare and quiescence PSDs are plotted in black and blue, respectively, and the magnetar PSD is shown in orange.    

F1 and F2 exhibit broad features at $f<10^{-3}$~Hz attributed to the flare envelopes as discussed in Section 5 and shown in Figure 2. In contrast, at higher frequencies ($f>10^{-2}$~Hz) both flares drop in power such that the flare PSDs overlap with or are within the uncertainties of the \sgra\ quiescent and magentar data.  Behavior in this regime is thus not unique to the flares, and can be attributed to noise.  Within the frequency range ($10^{-3}<f<10^{-2}$~Hz), however, the flare PSDs show hints of several small peaks that are not as visible in the quiescent or magnetar power spectra. These small-scale features have low significance and may be due to the instrumental effects discussed in Section \ref{sec:badpix} or harmonics of the structure from the flare envelope. Note that the $\sim3$s ($\sim3\times10^{-1}$~Hz) period of the magnetar would not be visible on the frequency range plotted in Fig. \ref{fig:appsd_1}.  

\begin{figure}
\center{
\includegraphics[scale=0.4]{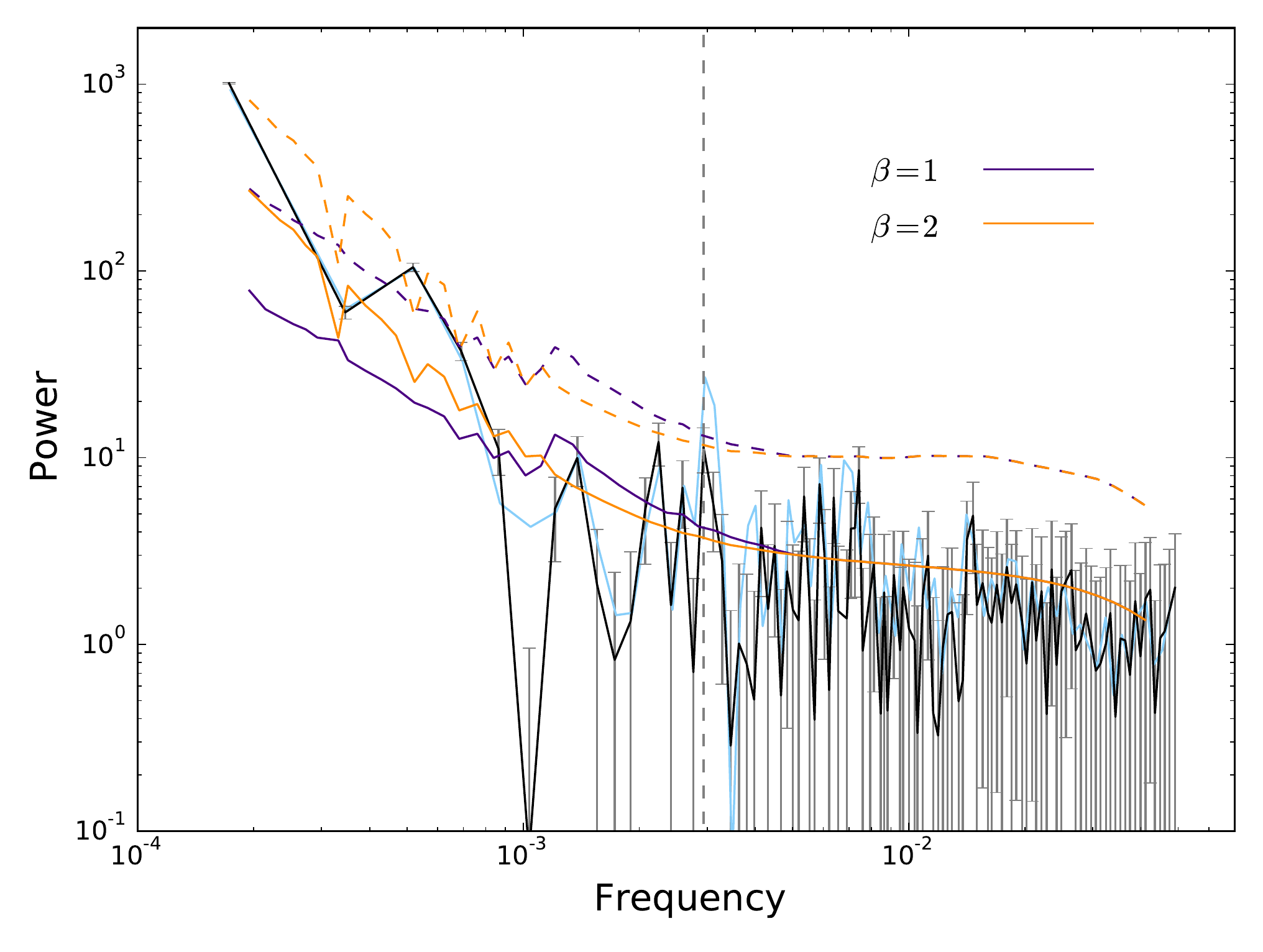}
\includegraphics[scale=0.4]{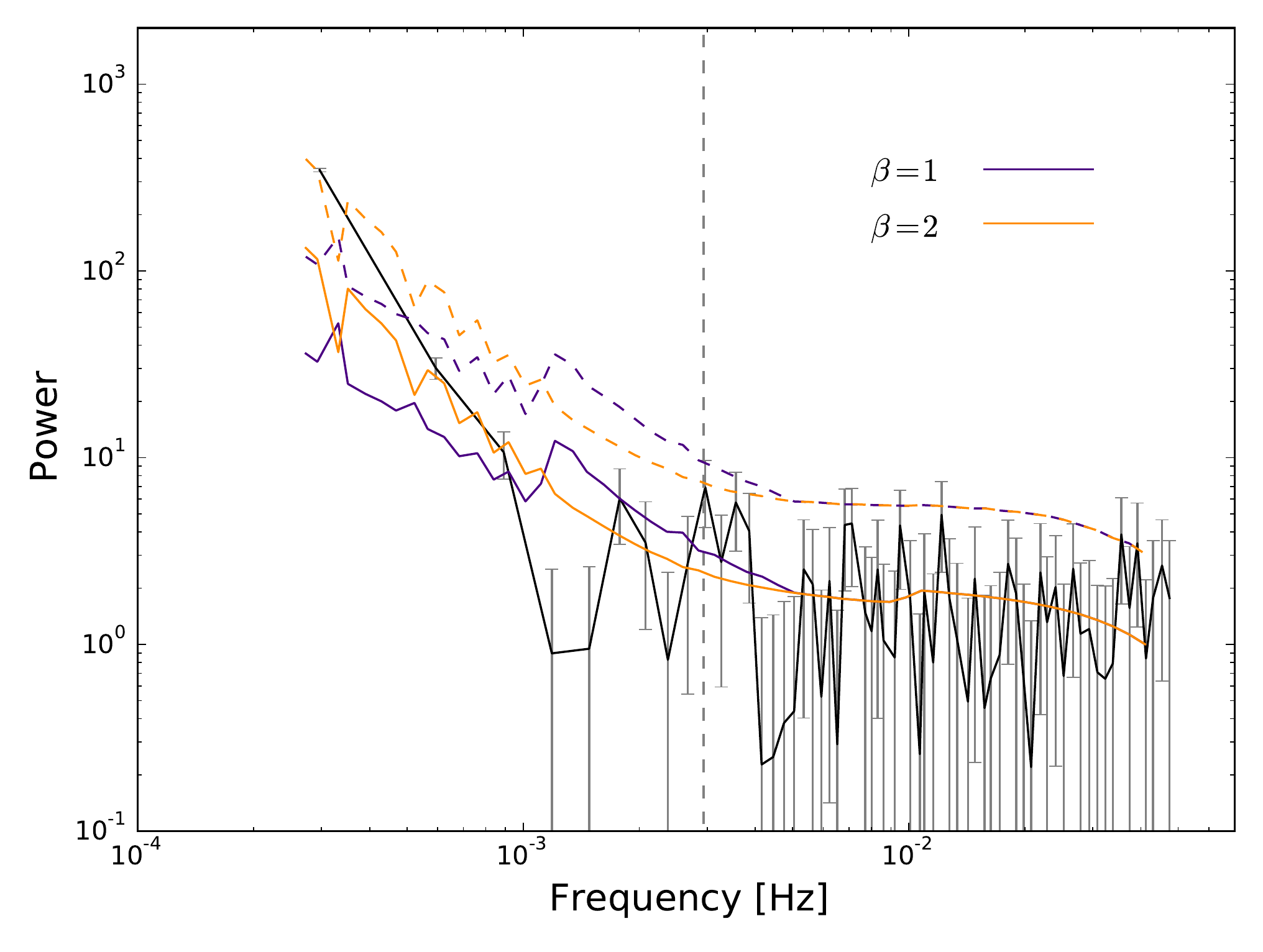}
}
\figcaption{
Leahy-normalized power spectral densities, derived via a fast Fourier transform (FFT), for the bright flare F1 (\textbf{\it left}\,) and flare F2 (\textbf{\it right}\,). The pale blue curve for F1 shows the PSD from data obtained with standard \Chandra\ data processing, which includes instrumental effects.  The spurious QPO at $\nu=2.99$~mHz in flare F1 is indicated by the dashed gray line, it is shown for flare F2 on the right for comparison only. The black curve shows the PSD with proper data filtering (\S\ref{sec:badpix}). For both flares, errors with $\beta=1.0$ and $\beta=2.0$ are shown in dark purple and dark yellow, respectively. The solid lines indicate $50\%$ contours, while the dashed represent $90\%$ contours.  
}
\label{fig:appsd_beta}
\end{figure}

We probe these midfrequency features by exploring two independent methods of quantifying the significance of the PSD, discussed in detail in Sections \ref{sec:errors1} and \ref{sec:errors2}. For the latter, we perform Monte Carlo simulations of the flare light curve, and for the former we generate PSDs under the worst-case assumption that all the flare photons arise from a combination of red and white noise. Figure \ref{fig:lcs} shows the $50\%$ and $90\%$ confidence intervals where the red noise is defined by a slope of $\beta=1.0$.  The spurious $\sim3$~mHz QPO is the only feature aside from the flare envelope that rises above the noise curve.  

However, a range of red noise slopes are possible in \sgra\ (see Section 5.1) and we therefore investigate the significance of the power spectral features where $\beta=2.0$ represents the red noise. Figure \ref{fig:appsd_beta} shows the PSD for the flare F1 with the standard \Chandra\ data processing in pale blue and the updated filtering defined in Section \ref{sec:badpix} in black. While the two curves are very similar, the QPO resulting from instrumentation effects is clearly seen in the blue PSD.  Overlaid are the error curves from the simulations where the red noise is defined with slopes of $\beta=1.0$ and $\beta=2.0$ in dark purple and dark yellow, respectively.  The $50\%$ confidence intervals are represented by a solid line, while the $90\%$ confidence limits (c.l.) are dashed.  

The white noise $\beta=0$ is dominant at higher frequencies, while the red noise component is relevant in the lower frequency regime.  This is evident as the curves with $\beta=1.0$ and $\beta=2.0$ diverge at $f<10^{-3}$~Hz.  In both scenarios there are a few features that rise above the solid $50\%$ c.l., but none that rise above the dashed 90\% confidence intervals and we conclude that no high frequency signal is detected in these bright flares. 
 
\section{Lightcurve Models}
\setcounter{figure}{0}
\setcounter{table}{0}
\renewcommand{\thetable}{B.\arabic{table}}
\label{app:b}

\begin{figure}[t]
\center{
\includegraphics[trim={0.5cm 1cm 1.5cm 1cm},clip,scale=0.65]{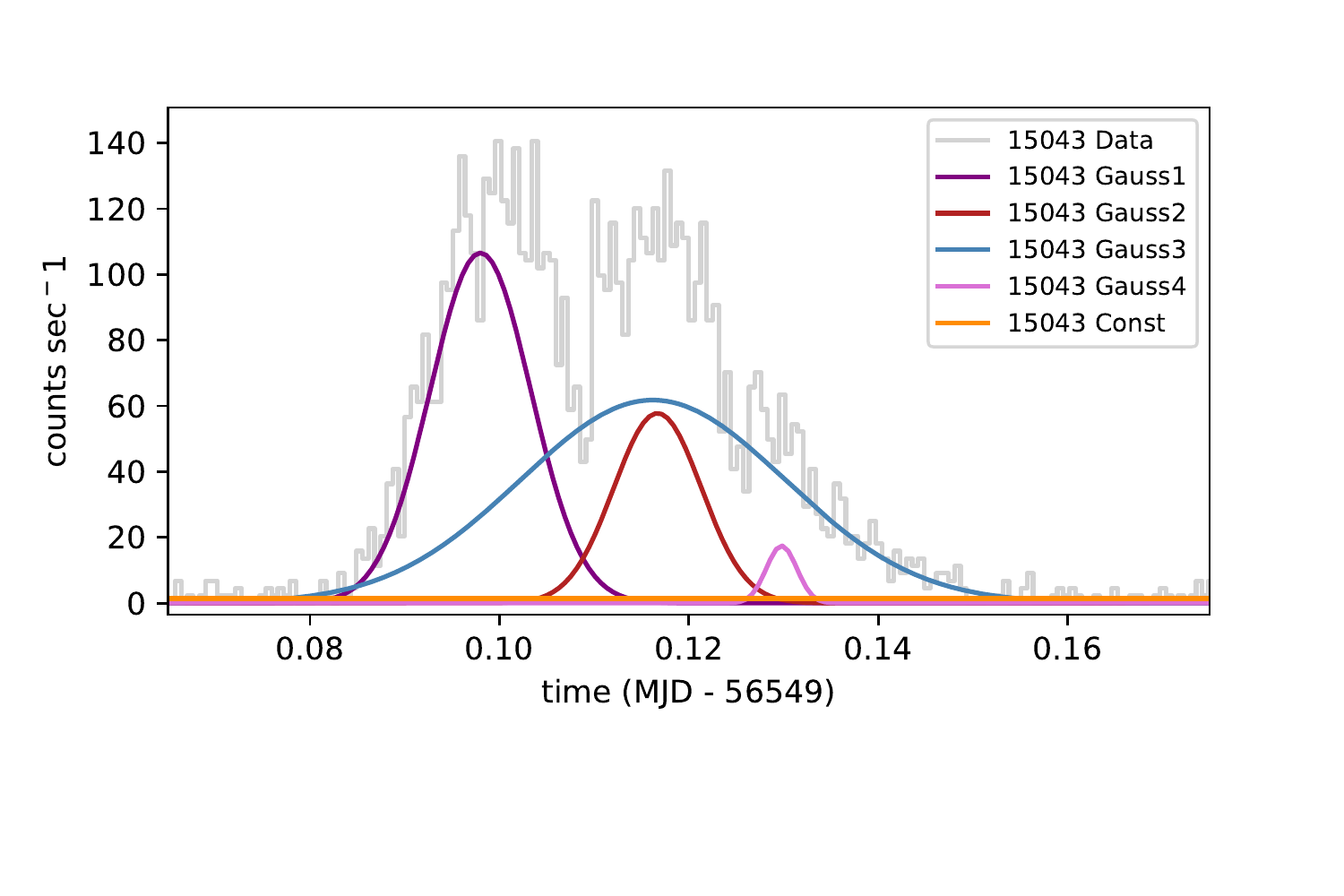}
\includegraphics[trim={1cm 1cm 1.5cm 1cm},clip,scale=0.65]{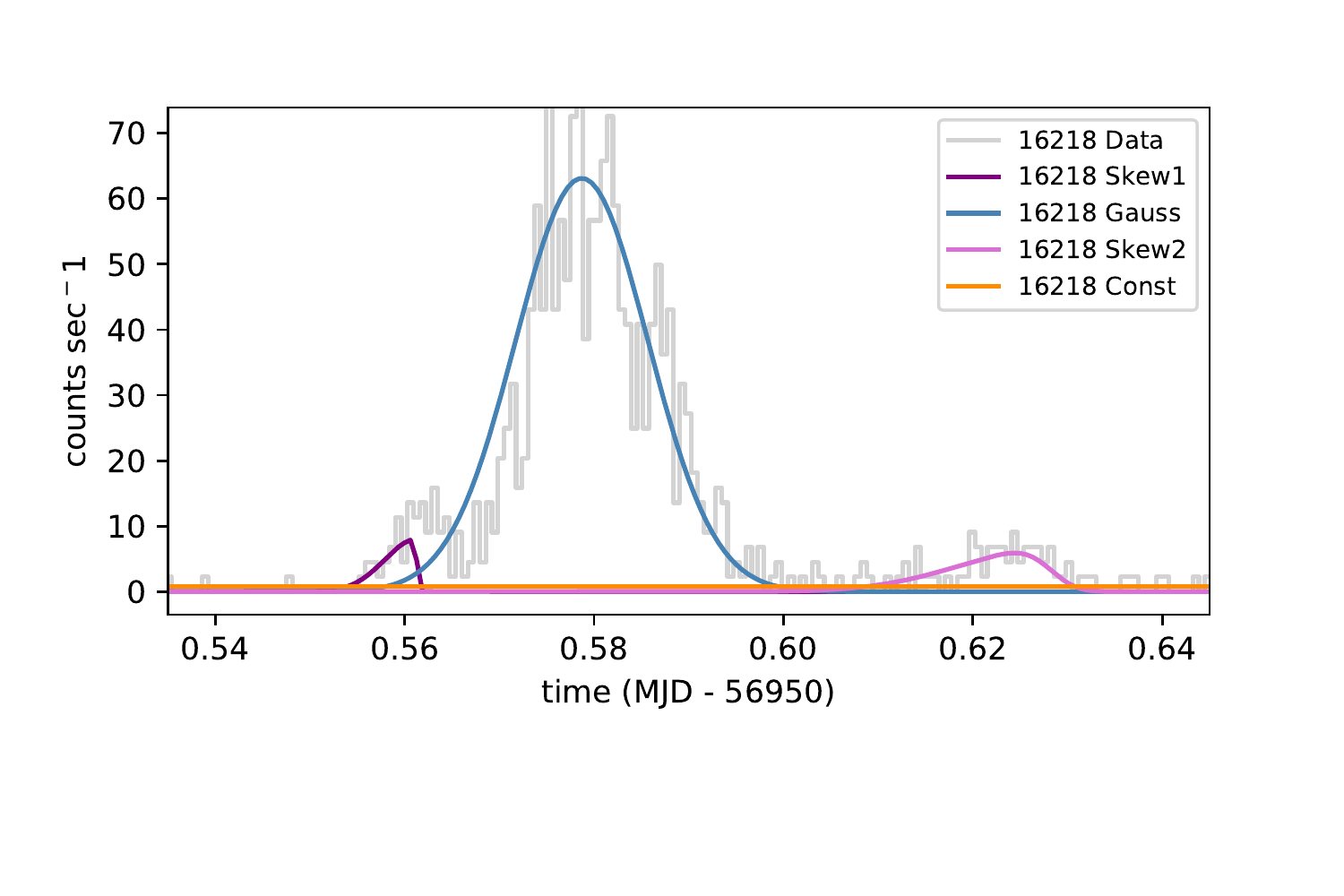}
}
\figcaption{Model components for the F1 and F2 lightcurve fits from Equations \ref{eqn:gauss} and \ref{eqn:sgauss} and parameters in Table \ref{tab:lcmodels}. The composite models are plotted as red curves in the top panels of Figure \ref{fig:lcs} and in all four panels of Figure \ref{fig:alcs}. The \Chandra\ X-ray data are shown in light gray, with $\sim$100 s binning (see \S\ref{sec:errors2}).}
\label{fig:appsd_2}
\end{figure}

The models fit to the X-ray lightcurves for F1 and F2, and used in the analyses above, are composed of a sum of four Gaussians plus a constant for F1, and a sum of two Gaussians and two skewed Gaussians plus a constant for F2. These take the standard functional forms  
\begin{equation}
    g(t; A, \mu, \sigma) = \frac{A}{{\sigma \sqrt {2\pi } }}e^{{{ - \left( {t - \mu } \right)^2 } \mathord{\left/ {\vphantom {{ - \left( {t - \mu } \right)^2 } {2\sigma ^2 }}} \right. \kern-\nulldelimiterspace} {2\sigma ^2 }}}
\label{eqn:gauss}
\end{equation}
\begin{equation}
    f(t; A,\mu,\sigma,\gamma) = \frac{A}{{\sigma \sqrt {2\pi } }}e^{{{ - \left( {t - \mu } \right)^2 } \mathord{\left/ {\vphantom {{ - \left( {t - \mu } \right)^2 } {2\sigma ^2 }}} \right. \kern-\nulldelimiterspace} {2\sigma ^2 }}} \left\{ 1 + {\rm erf}\left[ \frac{\gamma (t - \mu )}{\sigma \sqrt {2 } } \right] \right\}
\label{eqn:sgauss}
\end{equation}

where the parameters are amplitude (A), center ($\mu$), characteristic width ($\sigma$), and gamma ($\gamma$), and {\tt erf}() is the error function.

\begin{deluxetable}{lcccc}
\tabletypesize{\footnotesize}
\tablewidth{0pt}
\tablecaption{Lightcurve Models}
\tablehead{{Function} & {A (cnts/s)} & {$\mu$ (MJD)} & {$\sigma$ (MJD)} & {$\gamma$}} 
\startdata
\multicolumn{5}{l}{{\bf 15043}}\\
\hline{}
Gaussian 1 & 0.0114 & 56549.098 & 0.0053 & \ldots \\
Gaussian 2 & 0.0173 & 56549.116 & 0.0140 & \ldots \\
Gaussian 3 & 0.0054 & 56549.117 & 0.0046 & \ldots \\
Gaussian 4 & 0.0006 & 56549.130 & 0.0016 & \ldots \\
Constant   & 0.0112 & \ldots & \ldots & \ldots \\
\multicolumn{5}{l}{{\bf 16218}}\\
\hline{}
Skew Gauss 1 & 0.0003 & 56950.561 & 0.0034 & $-$15.950 \\
Gaussian       & 0.0089 & 56950.579 & 0.0070 & \ldots \\
Skew Gauss 2 & 0.0006 & 56950.628 & 0.0094 & $-$4.278 \\
Constant       & 0.0066 & \ldots & \ldots & \ldots \\
\enddata
\tablenotetext{}{The lightcurve models for F1 and F2 are described by the parameters above, using a combination of Equations \ref{eqn:gauss} and \ref{eqn:sgauss}. Each component is shown separately in Fig. \ref{fig:appsd_2} and the composite models are plotted as red curves in the top panels of Fig. \ref{fig:lcs} and in all four panels of Fig. \ref{fig:alcs}.}
\label{tab:lcmodels}
\end{deluxetable}

We plot these components in Figure \ref{fig:appsd_2} and give the detailed fits in Table \ref{tab:lcmodels} for reference. The composite lightcurve models for F1 and F2 are shown as solid red lines in the top panels of Fig. \ref{fig:lcs} and in all four panels of Fig. \ref{fig:alcs}.

\newpage
\bibliographystyle{apj}
\bibliography{bibliography}

\end{document}